\documentclass[a4paper,11pt]{article}
\pdfoutput=1 

\usepackage{jcappub} 

\usepackage[T1]{fontenc} 
\usepackage[separate-uncertainty]{siunitx}[=v2]
\usepackage[style=nature]{biblatex}
\usepackage{indentfirst}
\usepackage{graphicx}
\usepackage{subcaption}
\usepackage{float}
\usepackage[normalem]{ulem} 
\DeclareSIUnit \parsec {pc}
\usepackage{lineno}
\usepackage{tablefootnote}
\usepackage{multirow}
\usepackage{threeparttable}

\usepackage{subcaption}
\usepackage{orcidlink}

\usepackage[symbol]{footmisc}

\abstract{

Core-collapse supernovae mark the end of life of massive stars. However, despite their importance in astrophysics, their underlying mechanisms remain unclear. Neutrinos that emerge from the dense core of the star offer a promising way to study supernova dynamics. A strategy is presented to improve the potential of the KM3NeT neutrino telescope to detect core-collapse supernovae in our Galaxy or the Large Magellanic Cloud by further exploiting the properties of its optical modules equipped with multiple photomultipliers. A supernova burst is expected to produce a sudden hit rate increase in the KM3NeT detectors
, which could be used to detect a supernova even in the absence of triggers from other experiments. New observables have been defined for individual optical modules that exploit the geometry and time distribution of the detected hits, enabling a better discrimination between signal and background signatures. In addition, a thorough investigation of the related systematic uncertainties is presented for the first time. When implemented, this new methodology allowed KM3NeT to probe $46$\% more Galactic core-collapse supernova candidates than with the previous trigger strategy, reaching the dense Galactic bulge. It is now expected that, once completed, KM3NeT will achieve full Galactic sensitivity to core-collapse supernovae independently from other experiments.
}

\begin{document}

\title{Optimizing the potential of KM3NeT in detecting core-collapse supernovae}





\author[b,a]{O.~Adriani\,\orcidlink{0000-0002-3592-0654}}
\author[c,be]{A.~Albert}
\author[d]{A.\,R.~Alhebsi\,\orcidlink{0009-0002-7320-7638}}
\author[d]{S.~Alshalloudi}
\author[e]{M.~Alshamsi}
\author[f]{S. Alves Garre\,\orcidlink{0000-0003-1893-0858}}
\author[g]{F.~Ameli}
\author[h]{M.~Andre}
\author[i]{L.~Aphecetche\,\orcidlink{0000-0001-7662-3878}}
\author[j]{M. Ardid\,\orcidlink{0000-0002-3199-594X}}
% ...
\affiliation[a]{Universit{\`a} di Firenze, Dipartimento di Fisica e Astronomia, via Sansone 1, Sesto Fiorentino, 50019 Italy}
\affiliation[b]{INFN, Sezione di Firenze, via Sansone 1, Sesto Fiorentino, 50019 Italy}
\affiliation[c]{Universit{\'e}~de~Strasbourg,~CNRS,~IPHC~UMR~7178,~F-67000~Strasbourg,~France}
\affiliation[d]{Khalifa University of Science and Technology, Department of Physics, PO Box 127788, Abu Dhabi,   United Arab Emirates}
\affiliation[e]{Aix~Marseille~Univ,~CNRS/IN2P3,~CPPM,~Marseille,~France}
\affiliation[f]{IFIC - Instituto de F{\'\i}sica Corpuscular (CSIC - Universitat de Val{\`e}ncia), c/Catedr{\'a}tico Jos{\'e} Beltr{\'a}n, 2, 46980 Paterna, Valencia, Spain}
\affiliation[g]{INFN, Sezione di Roma, Piazzale Aldo Moro, 2 - c/o Dipartimento di Fisica, Edificio, G.Marconi, Roma, 00185 Italy}
\affiliation[h]{Universitat Polit{\`e}cnica de Catalunya, Laboratori d'Aplicacions Bioac{\'u}stiques, Centre Tecnol{\`o}gic de Vilanova i la Geltr{\'u}, Avda. Rambla Exposici{\'o}, s/n, Vilanova i la Geltr{\'u}, 08800 Spain}
\affiliation[i]{Subatech, IMT Atlantique, IN2P3-CNRS, Nantes Universit{\'e}, 4 rue Alfred Kastler - La Chantrerie, Nantes, BP 20722 44307 France}
\affiliation[j]{Universitat Polit{\`e}cnica de Val{\`e}ncia, Instituto de Investigaci{\'o}n para la Gesti{\'o}n Integrada de las Zonas Costeras, C/ Paranimf, 1, Gandia, 46730 Spain}
% ...
\affiliation[be]{Universit{\'e} de Haute Alsace, rue des Fr{\`e}res Lumi{\`e}re, 68093 Mulhouse Cedex, France}
% ...
\emailAdd{km3net-pc@km3net.de}
\emailAdd{elhedri@apc.in2p3.fr}
\emailAdd{goos@apc.in2p3.fr}



\maketitle
\flushbottom

\newpage
\section{Introduction}
\label{sec:intro}

Hot compact objects such as core-collapse supernovae (CCSNe) and binary neutron-star mergers are characterized by the thermal emission of MeV-scale neutrinos. The first and only observation of such neutrinos occurred in association with the SN1987A core-collapse supernova in the Large Magellanic Cloud~\cite{walker1987making}. This observation confirmed that around $99\%$ of the gravitational binding energy of a collapsing star is released in the form of neutrinos~\cite{schramm1990new}, supporting the hypothesis that neutrino-driven mechanisms power supernova explosions. A total of only about $24$ neutrinos were observed across the  Kamiokande~\cite{hirata1988observation}, IMB~\cite{bionta1987observation}, and Baksan~\cite{alekseev1987possible} detectors, leaving significant gaps in our understanding. The mechanisms by which neutrino heating in the stalled shock wave leads to a successful explosion and asymmetric neutrino emission contributes to the explosion, remain subjects of active research~\cite{suzuki2024neutrinos}. To address these and other open questions~\cite{horiuchi2018can}, detecting the next CCSN neutrino emission is essential. The CCSN neutrino signal precedes electromagnetic emissions from a supernova. Consequently, a fast combination of measurements from neutrino telescopes around the globe to triangulate the source's position~\cite{al2021snews, coleiro2020combining} would allow for alerting optical telescopes to the upcoming visible counterpart. Such coordinated efforts between neutrino observatories and optical telescopes promise to provide a better understanding of these cataclysmic phenomena.

KM3NeT is a network of large-scale underwater neutrino telescopes currently under construction in the Mediterranean Sea~\cite{KM3Net_LoI}. It comprises two detectors: ARCA, off-shore Capo Passero, Sicily, Italy, at a depth of $3.5$~km and ORCA, off the coast of Toulon, France, at a depth of $2.5$~km. They are respectively optimised for the observation of astrophysical neutrinos in the TeV--PeV energy range and for the study of atmospheric neutrinos with energies between $1$--$100$~GeV. Both detectors consist of a 3-dimensional array of Digital Optical Modules (DOMs)~\cite{aiello2022km3net} that detect the Cherenkov light along the path of charged particles emerging from a neutrino interaction. These DOMs are arranged along vertical string-like detection units (DUs) which are anchored to the seafloor. Each DOM is equipped with $31$ photomultiplier tubes (PMTs) uniformly distributed in almost all directions.
Although thermal supernova neutrinos have energies of the order of only \SI{10}{\mega\eV}, i.e. far below the nominal sensitivity range of KM3NeT, the combination of the large instrumented volumes of KM3NeT with the multi-PMT design of the DOM makes the detection and characterization of a CCSN possible. In \cite{van2021km3net} and \cite{aiello2022implementation}, new methods to record the CCSN neutrino luminosity as a function of time, constrain the CCSN neutrino energy spectrum, and detect the signature of hydrodynamical instabilities were presented. Moreover, these articles proposed a strategy
to identify CCSN neutrinos without relying on information from other experiments
, exploiting the fact that $\mathcal{O}(\SI{10}{\mega\eV})$ neutrino interactions will lead to the observation of multiple PMT hits within a single DOM in a narrow, $\mathcal{O}(\SI{10}{\nano\s})$, time window. A supernova burst could thus be detected by observing an increase of such hit coincidences above the expected background level. Using this strategy, the probability for the \emph{completed} ARCA and ORCA detectors, combined, to identify the next Galactic core-collapse supernova independently from other experiments will be higher than $95\%$. However, this detection probability is significantly lower for the presently-deployed partial detector configurations. Assessing and improving the CCSN detection capabilities of these \emph{partial} configurations is therefore essential.

In this paper, a new strategy is introduced, which improves the identification of low-energy signals in KM3NeT by characterizing the associated multi-hit coincidences using machine learning techniques. This strategy is used to provide an accurate estimate of low-energy backgrounds in KM3NeT, thus enhancing the detectability of CCSN neutrinos. A description of the KM3NeT detectors is given in Section~\ref{sec:km3net}. In Section~\ref{sec:CCSNtriggers}, the search for CCSN neutrinos with KM3NeT is presented and the set of new observables exploiting the structure of the KM3NeT DOMs is introduced. Section~\ref{sec:km3bg} describes how these new observables can be used to characterize the background signatures produced in KM3NeT by radioactive decays in the seawater and atmospheric muons. The new strategy based on machine learning to detect CCSN neutrino interactions is detailed in Section~\ref{sec:filtering}. The resulting sensitivity of KM3NeT to a CCSN neutrino burst is presented in Section~\ref{sec:sensitivity}. Finally, concluding remarks are given in Section~\ref{sec:conclusion}.

\section{The KM3NeT infrastructure}
\label{sec:km3net}


While both KM3NeT detectors rely on the same detection principle, the density of instrumentation used in ORCA and ARCA is optimised for their respective science objectives. All KM3NeT DUs support 18 DOMs, whose spacing is dictated by the energy range of interest for each detector.  In ARCA, the vertical DOM spacing is $\sim 36$~m and the horizontal DU spacing is of $\sim 90$~m. In ORCA, the vertical DOM spacing is $\sim 9$~m and the horizontal DU spacing of $\sim 20$~m. When completed, ARCA and ORCA will be composed of $230$ and $115$ DUs, respectively, grouped in building blocks of $115$ DUs. Each building block will hence contain $2070$ DOMs. At the time of writing, $51$ DUs have been installed for ARCA and $28$ DUs for ORCA. In the following, partial configurations of ARCA and ORCA are referred to by the detector name followed by the number of deployed DUs at the time. For example, the current configurations are named ARCA51 and ORCA28.



A significant challenge for a seawater detector like KM3NeT is the presence of large environmental backgrounds due to living organisms (bioluminescence) and ambient radioactivity. Characterizing and monitoring these backgrounds has been one of the main motivations for the multi-PMT design of the KM3NeT DOM~\cite{aiello2022km3net}. A DOM consists of $31$ three-inch photomultiplier tubes, enclosed in a pressure-resistant glass sphere of radius \SI{22}{\centi\m}. Each photomultiplier measures the time of detection of incoming photons with nanosecond accuracy. 

The multi-PMT structure of the DOM leads to an improved reconstruction of high-energy neutrino signatures, by providing sensitivity to the direction of incoming photons. The design also enhances the KM3NeT sensitivity to MeV-scale charged particles. Indeed, these particles induce the emission of Cherenkov light over distances of $\mathcal{O}(\SI{10}{\centi\m})$, potentially resulting in multiple PMT hits within a single DOM. 


The data acquisition system of KM3NeT is based on the ``all-data-to-shore'' concept~\cite{KM3Net_LoI}. The data stream, consisting of all PMT hits above a threshold of $0.3$ photoelectrons, is subdivided into \SI{100}{\milli\s} segments called \emph{timeslices}. The baseline hit rate within these timeslices is about $ \SI{7}{\kilo\Hz}$ per PMT, mainly due to radioactivity from $^{40}$K in water and to bioluminescence. Bioluminescence manifests itself as a superposition of sudden intense bursts of light above a slowly-varying seasonal component~\cite{priede2008potential}. To reduce the amount of these bursts, PMTs with a data-taking rate exceeding \SI{20}{\kilo\Hz} in a given timeslice are deactivated for the entire timeslice duration. This high-rate veto affects the performance of CCSN searches~\cite{aiello2022implementation}. In this article, following the convention from~\cite{aiello2022implementation}, the impact of the high-rate-veto is assessed using the fraction of active PMTs, $f_A$, computed over all working DOMs. As in~\cite{van2021km3net}, the capability of KM3NeT to detect CCSNe is evaluated for timeslices with $f_A > 99$\%. The impact of $f_A$ on the performance of KM3NeT is discussed in Section~\ref{sec:sensitivity}.

In the primary data stream, the hits stored in timeslices are used to identify \emph{triggered events}, defined as groups of causally-connected hit coincidences distributed over multiple DOMs and corresponding to specific topologies. These triggers are used to identify and characterize high-energy neutrinos and atmospheric muons~\cite{KM3NeT:2024paj}. For CCSN neutrino studies, multi-hit coincidences within single DOMs are saved in a separate data stream called a \emph{SN timeslice}. Such coincidences require at least 4 hits observed in a single DOM within a $15$~nanosecond window from PMTs whose axes lie within an angle of $90^\circ$.~\cite{aiello2022implementation}. With these conditions, contributions from bioluminescence in the SN timeslice stream are negligible. Thus, with triggered events and SN timeslices, KM3NeT monitors neutrino interactions over two length scales: the \emph{inter-DOM} scale, primarily sensitive to particles with energy in the GeV to PeV range, and the \emph{intra-DOM} scale, which allows for probing particles down to an energy of a few MeVs. 

\section{Searching for CCSN neutrinos}
\label{sec:CCSNtriggers}

A core-collapse supernova occurring  at a distance of \SI{10}{\kilo\parsec} would generate an intense burst of neutrinos, with fluxes potentially as high as $10^{10}$--$10^{11}$~cm$^{-2}$ within the first few hundred milliseconds of the collapse. The interactions of these neutrinos in water are expected to produce multi-hit coincidences in individual DOMs, which would be recorded in the SN timeslices introduced above. Thus, KM3NeT could detect a nearby supernova by identifying an excess of such coincidences above the expected background.

\subsection{Analysis overview}
\label{subsec:overview}

The analysis presented here aims at improving the performance of the search for CCSN neutrinos discussed in~\cite{van2021km3net} and \cite{aiello2022implementation}. It adopts the same definition of a \emph{single-DOM event} as a cluster of PMT hits observed in a DOM within a \SI{10}{\nano\s} time window. Single-DOM events are preselected by requiring that they belong to SN timeslices and pass a dedicated \emph{triggered event veto} which removes events associated with highly energetic particles (e.g.\ atmospheric muons) and PMT afterpulses~\cite{van2021km3net}. 

In the CCSN analysis of~\cite{van2021km3net}, the preselected single-DOM events were further filtered according to their \emph{multiplicity}, i.e. the number of hits they contain. 
The analysis proposed here increases the KM3NeT sensitivity to CCSN neutrinos by considering, for each single-DOM event, not only the number of PMT hits but also their position, timing, and charge. Since a DOM contains $31$ PMTs, accounting for the properties of all the hits in each event would significantly increase the complexity of the analysis. To address this challenge, a set of single-DOM observables has been defined that reflects the properties of individual hits.

\subsection{Single-DOM observables}
\label{subsec:definitions}

The observables defined in this section describe four properties of a single-DOM event: the position of the hit cluster on the DOM, the spatial and temporal correlations of the hits, and the intensity of the signal.

\begin{figure}[t!]
    \centering
    \includegraphics[width=0.35\linewidth]{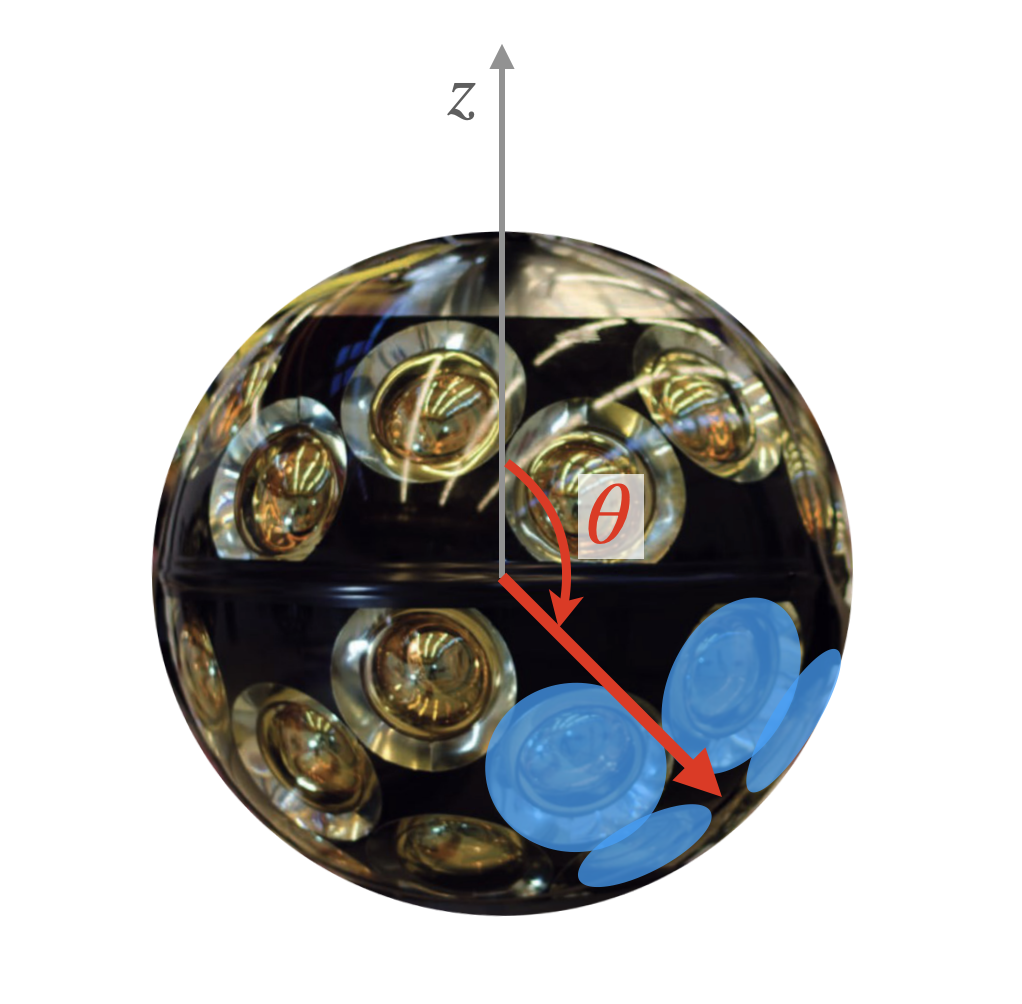}
    \caption{Picture representing a single-DOM event with multiplicity $4$, with the hit PMTs highlighted in blue. The $\vec{R}$ vector from Equation~\ref{eq:Rvec} is shown in red. Its magnitude $|R|$ is correlated with the spatial concentration of the PMT hits. In this particular case, since the PMT hits are close together, $|R|$, $\cos\alpha_{i,i+1}$ and $\cos\alpha_{i,i+2}$ will be close to $1$, while $\sigma(\cos\alpha_{i,i+1})$ and $\sigma(\cos\alpha_{i,i+2})$ will be close to $0$. Here, $\cos\theta \sim -0.7$ as the hits are located in the lower hemisphere of the DOM.}
    \label{fig:DOM}
\end{figure}

\begin{itemize}
\item Five observables are considered to describe the \textit{spatial distribution} of hit PMTs across the DOM, taking the center of the DOM as the origin of the Cartesian coordinate system: i) the mean cosine of the angle between the axes of the hit PMTs consecutive in time, $\langle\cos \alpha_{i, i+1}\rangle$; ii) its standard deviation, $\sigma(\cos \alpha_{i, i+1} )$; iii) the mean cosine of the angle between next-to-consecutive hit PMTs, $\langle\cos \alpha_{i, i+2}\rangle$; iv) its standard deviation, $\sigma( \cos \alpha_{i, i+2} )$; v) the hit concentration $|R|$, defined as the magnitude of the vector
\begin{align}
    \label{eq:Rvec}
    \vec{R} = \frac{1}{M}\sum_{i=1}^M \vec{d}_i,
\end{align}
where $M$ is the multiplicity and $\vec d_i$ are the directions of the axes of the hit PMTs. A value of $|R|$ close to 1 indicates a concentrated signal, with PMT hits occurring close together, whereas $|R|$ is close to $0$ if the hits are randomly distributed across the DOM. 

\item Four observables are used to investigate the \textit{temporal spread} of the signal: the mean and the standard deviation of the time differences between consecutive hits, $\langle\Delta t_{i, i+1}\rangle$ and $\sigma(\Delta t_{i, i+1})$; the average hit timing $\langle\Delta t\rangle = \langle t_i - t_0\rangle$, where $t_0$ is the time of the first hit, and the corresponding standard deviation $\sigma(\Delta t)$.

\item The \textit{position} of the hit cluster on the DOM is described using the zenith angle $\theta$ of the $\vec R$ vector defined in Equation~\ref{eq:Rvec} (see Figure~\ref{fig:DOM}). A value of $\theta=0 (\pi)$ corresponds to the North (South) Pole of the DOM. 

\item The \textit{signal intensity} is described by two observables: the multiplicity and the total time-over-threshold (ToT), defined as the sum of the time-over-thresholds of the PMT hits~\cite{ToT}. 

\end{itemize}

\subsection{Comparison of signal and background}
\label{subsec:SignalBkg}

The discriminating power of the single-DOM observables introduced above is investigated by comparing their distributions for CCSN signal and background events. Background events, mainly due to radioactive decays in the seawater and atmospheric muons, are modeled using ORCA6 data (see Appendix~\ref{subsubsec:samples}). While radioactive decay products generate localized signatures that activate a small cluster of PMTs on a single DOM, atmospheric muons can travel hundreds of meters, resulting in many PMT hits across multiple DOMs along their track. Signal events are generated by simulating single-DOM events weighted by the CCSN spectrum of interest (see Appendix~\ref{subsec:SignalSimu}). Only single-DOM events passing the preselection steps of the CCSN search, i.e. belonging to SN timeslices and passing the triggered event veto, are considered. Here and in the rest of this paper, unless indicated otherwise, only data-taking periods with a fraction of active PMTs $f_A$ larger than $99\%$ are considered.

\begin{figure}[t!]
    \centering  \includegraphics[width=0.6\linewidth]{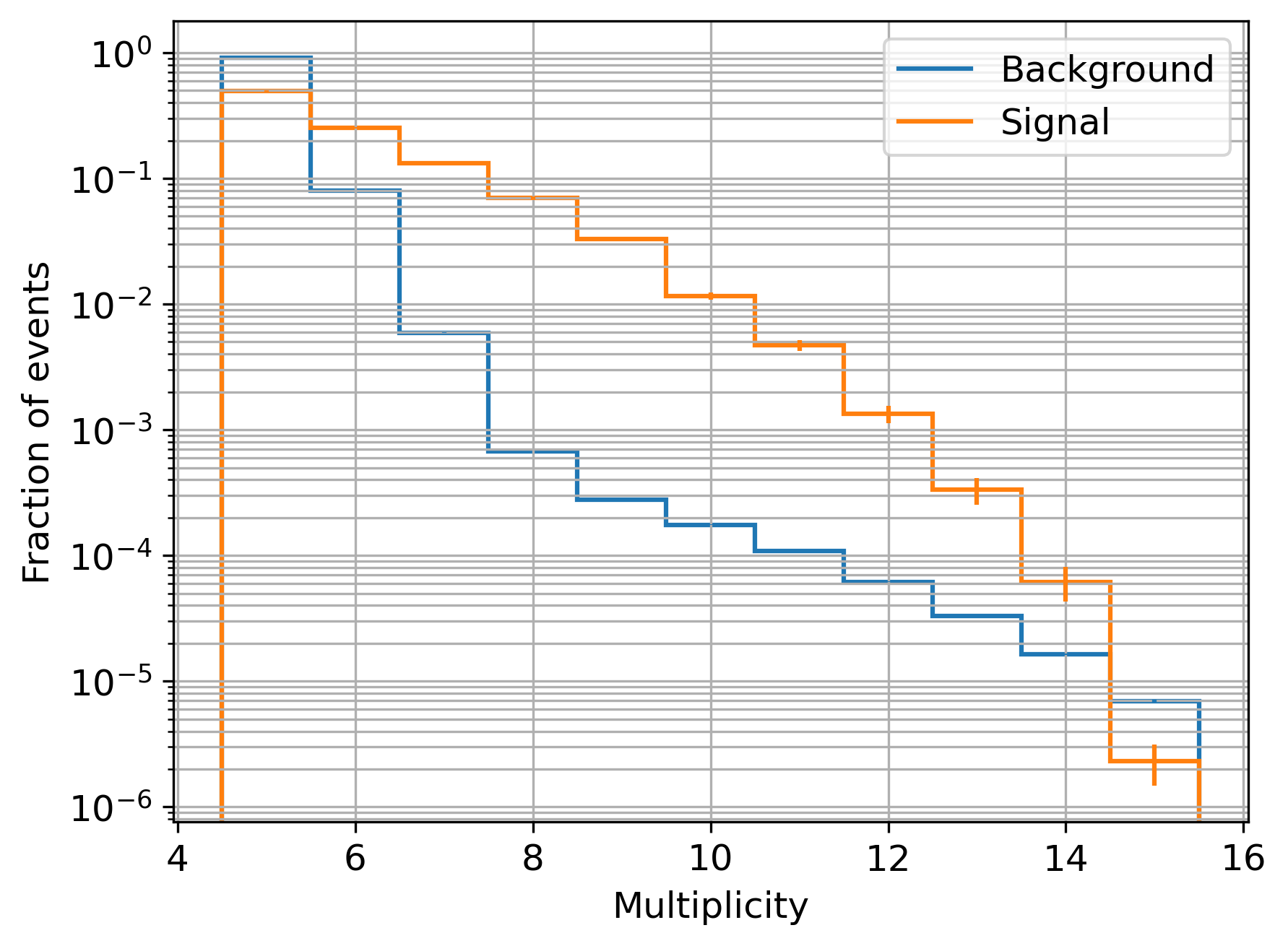}
    \caption{Multiplicity distributions for simulated CCSN signal events (orange) and background events from a $60$-day period of ORCA6 data (blue). The areas under the distributions are normalized. Only events passing the triggered event veto introduced in~\cite{van2021km3net} are included. For the signal, the $11~M_\odot$ progenitor from~\cite{tamborra2014neutrino} is used.}
    \label{fig:sig_data_mult}
\end{figure}

The multiplicity distributions for background and simulated signal events are shown in Figure~\ref{fig:sig_data_mult}. Intermediate multiplicities, between $6$ and $14$, are associated with the highest signal-to-background ratio. The shape of the background distribution, similar to a broken power law, reflects the background composition: radioactivity dominates up to multiplicity $7$ and atmospheric muons dominate above multiplicity $8$~\cite{goos2023searching}.

\begin{figure}[H]
    \centering
    \includegraphics[width=\linewidth]{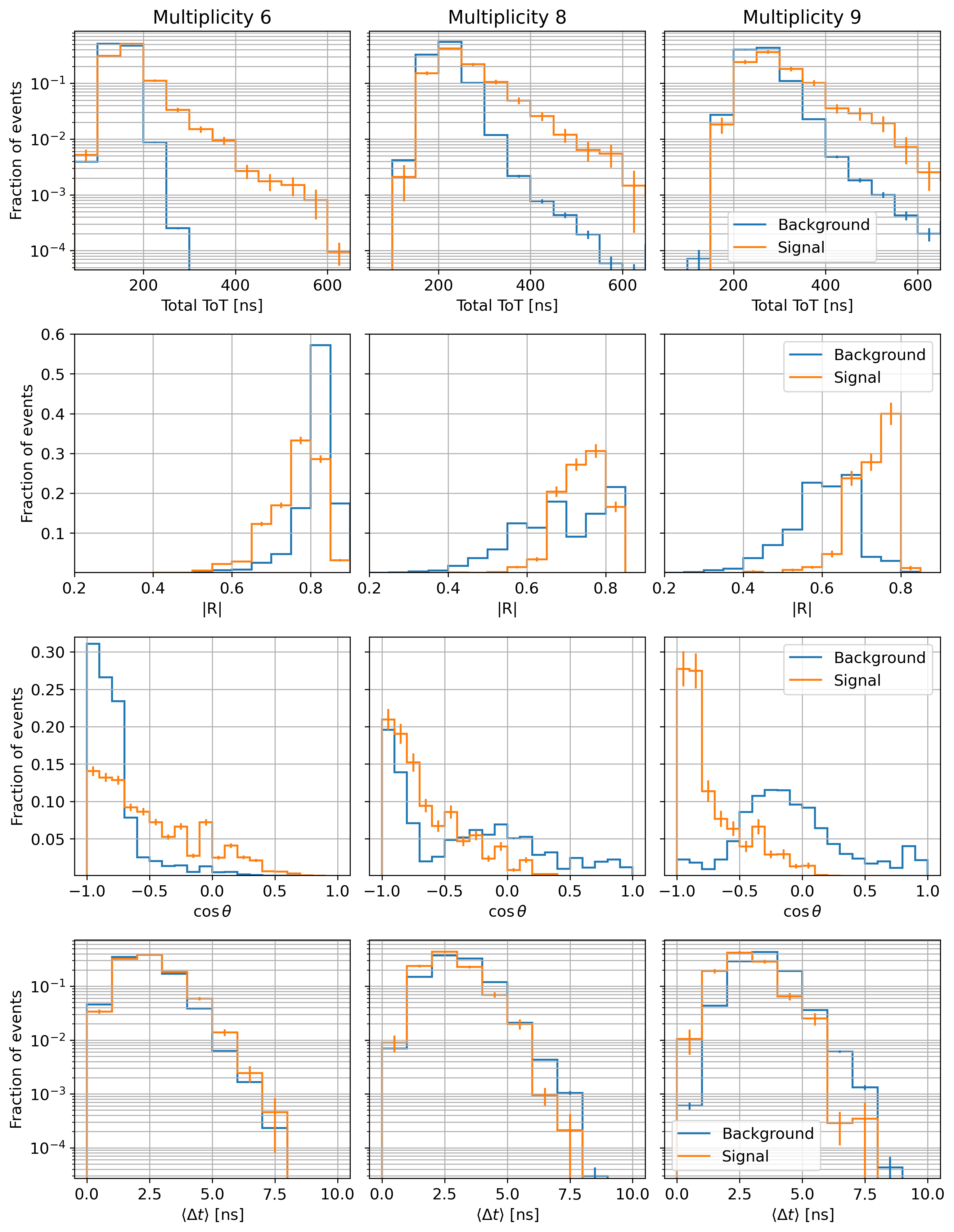}
    \caption{Distributions of the total ToT (first row), hit concentration $|R|$ (second row), $\cos\theta$ (third row), and $\langle\Delta t\rangle$ (fourth row) for signal (orange) and background (blue) considering only events with multiplicity $6$ (left), $8$ (center), and $9$ (right). The simulated signal is based on the $11 M_\odot$ model from~\cite{tamborra2014neutrino}. The background is taken from ORCA6 data and is composed of atmospheric muons and radioactive decays. The areas under all the distributions are normalized to unity. Only events that passed the triggered event veto are included.}
    \label{fig:observables}
\end{figure}

The distributions of total ToT, $|R|$, $\cos \theta$, and $\langle\Delta t\rangle$, for background and signal, are presented in Figure~\ref{fig:observables} for multiplicities $6$, $8$, and $9$, corresponding to radioactivity-dominated, mixed, and muon-dominated backgrounds, respectively. For a fixed multiplicity, MeV-scale neutrinos, radioactive decays, and atmospheric muons produce distinct hit patterns. In all cases, a larger fraction of events with high-ToT is expected for the signal compared to the background. This can be explained by the fact that CCSN neutrino interactions close to DOMs can lead to large ToTs, while high-ToT background events are mostly associated with triggered muons, which are largely removed by the triggered event veto. Furthermore, the $|R|$, $\cos\theta$, and $\langle\Delta t\rangle$  distributions at low multiplicity (radioactivity-dominated region) and high multiplicity (muon-dominated region) show opposite trends, with the CCSN signal occupying an intermediate position (see also~\cite{goos2023searching}). This behavior relates to the fact that hits from radioactive decays are concentrated in both time and space since $\sim$\SI{1}{\mega\eV} radioactive decay products travel short distances and must be produced close to a DOM to be detected. Moreover, radioactive decay signatures are typically observed near the bottom of the DOM since the lower DOM hemisphere is the most densely instrumented. Conversely, hits from atmospheric muons are dispersed over a large area and time interval, as muons travel long distances, and are mostly observed in the upper DOM hemisphere since atmospheric muons are downward-going. Finally, CCSN signal events have topologies similar to radioactivity events but are associated with more energetic particles. Hence, they often have intermediate $|R|,\langle\Delta t\rangle,$ and $\cos\theta$ values and can thus be distinguished from both background components. 

The distributions shown in Figure~\ref{fig:observables} highlight the potential of single-DOM observables to both separate radioactivity and atmospheric muons in KM3NeT data, and distinguish CCSN neutrino signatures from backgrounds. In this analysis, these capabilities are exploited to i) improve the characterization of low-energy backgrounds at KM3NeT and ii) expand KM3NeT's CCSN detection horizon for a wide range of supernova progenitors.
\\\newpage

\section{Background characterization at KM3NeT}
\label{sec:km3bg}

In this section, a data-driven method is presented to estimate the radioactivity and muon background rates. In the $6$--$10$ multiplicity range used in CCSN studies, radioactivity and muon backgrounds coexist and their relative proportions have, to date, been evaluated using simulations~\cite{radioactivity_simulation}. Measuring the muon and radioactivity rates separately for each multiplicity is essential to estimate the sensitivity of upcoming ARCA and ORCA configurations to CCSN neutrinos and assess the accuracy of the KM3NeT detector simulations. Beyond the CCSN analysis, estimating the radioactivity background rates over a wide range of multiplicities could enhance searches for other MeV-scale phenomena related to e.g. beyond-the-Standard-Model physics, and improve the modeling of high-energy neutrino signatures in ARCA and ORCA.

Although the total background rates in individual DOMs can be directly measured for all multiplicities~\cite{dom_rates}, determining the relative fractions of atmospheric muon and radioactivity events for a given multiplicity can be challenging. This is especially true in the CCSN multiplicity search region, where both backgrounds are present in comparable amounts once the event preselection described in Section~\ref{sec:CCSNtriggers} is applied. However, radioactivity and atmospheric muons leave distinctive hit patterns on individual DOMs, which can be distinguished using the single-DOM observables introduced in Section~\ref{sec:CCSNtriggers}. 

\subsection{Background properties}
\label{subsec:totalbg}
\begin{figure}
    \centering
    \includegraphics[width=\linewidth]{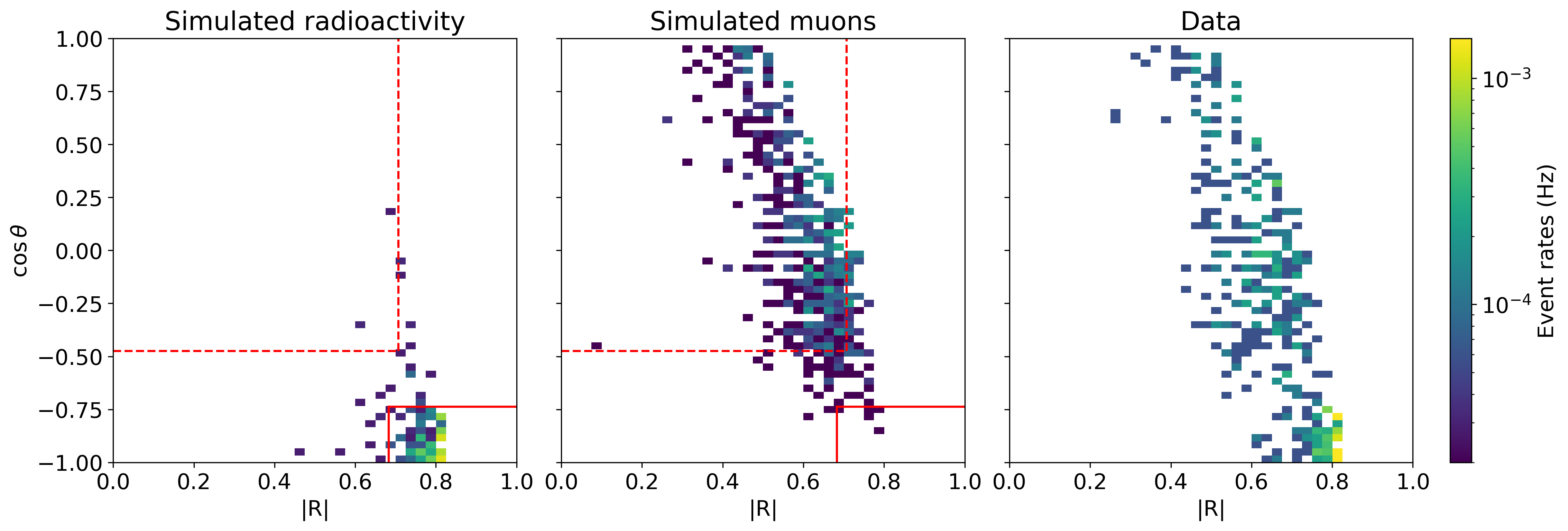}
    \caption{Two-dimensional $(|R|, \cos\theta)$ distributions for data and for simulated ORCA6 background events of multiplicity $8$ passing the CCSN preselection. Radioactivity and muon events, simulated as described in Appendix~\ref{sec:bkgmodels}, are shown in the left and middle panels while the data are shown in the right panel. In the region on the left of the dashed red lines (on the right of the solid red lines) $99\%$ of the background is due to atmospheric muons (radioactivity).}
    \label{fig:muons_r_ctheta}
\end{figure}

As shown in Figure~\ref{fig:observables}, the maximum separation between radioactivity and atmospheric muons is obtained for the hit concentration $|R|$ and $\cos\theta$, the cosine of the zenith angle of $\vec R$. In particular, at multiplicity~$8$, where radioactivity and atmospheric muon contributions are comparable, the $|R|$ and $\cos\theta$ distributions are bimodal with the signal occupying an intermediate position between the two modes. To further characterize these distributions for each background component, muon and radioactivity backgrounds are modeled using the KM3NeT simulation software described in Appendix~\ref{sec:bkgmodels}. The two-dimensional ($|R|$, $\cos\theta$) distributions for simulated radioactivity events, simulated muon events, and data, with multiplicity $8$, are shown in Figure~\ref{fig:muons_r_ctheta} for the ORCA6 configuration. As anticipated in Section~\ref{subsec:SignalBkg}, muon and radioactivity events occupy separate corners of the ($|R|$, $\cos\theta$) space. Hence, using these two observables is sufficient to identify muons and radioactivity single-DOM events. 

\subsection{Estimating radioactivity and muon background rates}
\label{subsec:rescale}

To estimate radioactivity and muon rates for a given dataset and multiplicity, rectangular control regions containing a fixed, large, fraction of either radioactivity or atmospheric muons are defined in the $(|R|,\cos\theta)$ space based on the respective background simulations. The radioactivity and atmospheric muon rates in data are then obtained by rescaling the background simulations to match the number of observed events in each region. Thus, each background simulation is associated with a multiplicity-dependent \emph{scaling factor}. In this study, to account for potential mismodeling of the background composition in the control regions, multiple scaling factors are derived for each background component and multiplicity, by defining control regions with dominant background fractions ranging from $90$\% to $99$\%. The mean of these scaling factors is used to estimate the background rate, while the total variation is taken as an uncertainty.

The radioactivity and muon rates have been estimated using the method described above for different data samples. For muons, several runs of ORCA6, ORCA15, and ARCA21 have been considered , without imposing any triggered event veto. For all the data samples, the size of the control regions increases with the multiplicity: in the $6$--$10$ multiplicity range, the maximal $|R|$ and minimal $\cos\theta$ values range from $0.43$ to $0.70$ and $0.90$ to $-1.0$, respectively.  For radioactivity, only one dataset was considered since the background rates depend only weakly on the detector geometry ---as radioactive decay signatures are associated to single DOMs. This dataset was built by combining all the good-quality runs of ORCA6 (see Appendix~\ref{subsubsec:samples}). This configuration was chosen due to i) the long data-taking period of more than 1 year, ii) the large fraction of events with a high fraction of active PMTs $f_A$, iii) the superior performance of the triggered event veto of ORCA compared to that of ARCA. To allow identifying radioactivity up to high multiplicities, the triggered event veto introduced in Section~\ref{sec:CCSNtriggers} has been applied. Here, the control regions shrink as the multiplicity increases. For the $5$--$10$ multiplicity range the minimal $|R|$ and maximal $\cos\theta$ values lie in $[0.68,0.8]$ and $[-0.96,-0.26]$, respectively.

\begin{figure}
    \centering
    \includegraphics[width=0.6\linewidth]{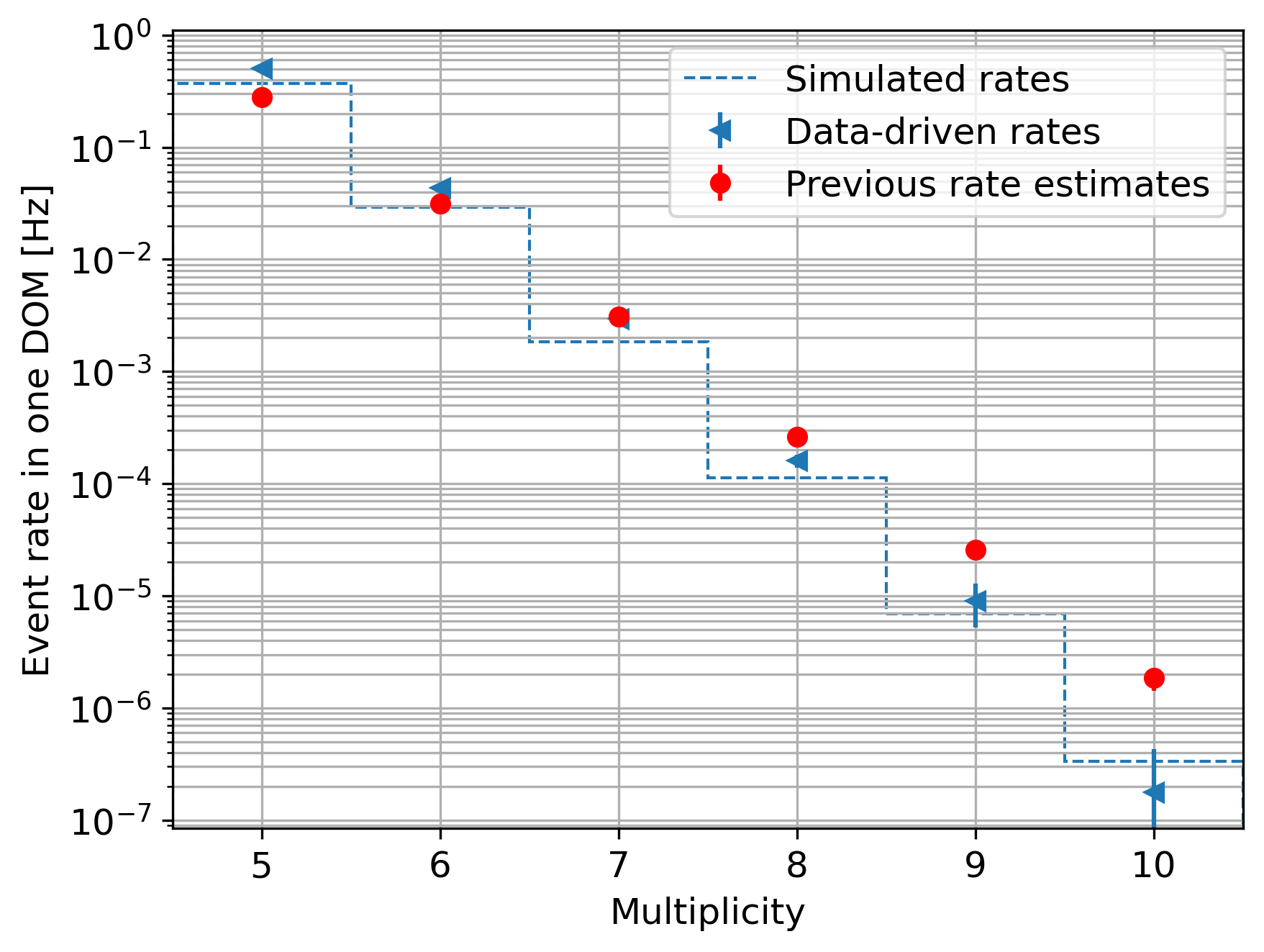}
    \caption{Radioactivity rates predicted by the simulations described in Appendix~\ref{subsec:radioactivity} (dashed line) and rescaled for each multiplicity using the procedure described in Section~\ref{subsec:rescale} (blue triangles). The uncertainty bars on the rescaled rates account for both the finite statistics in the control regions and the modeling uncertainties on the background fractions. The red dots indicate the radioactivity rate estimations obtained using simulations tools, as in the previous CCSN analysis~\cite{van2021km3net}.}
    \label{fig:compare_radioactivity_rates}
\end{figure}

For all tested samples, the muon rates do not vary significantly with the multiplicity and remain within $15\%$ of the simulation predictions. Conversely, the radioactivity event rates evaluated using this data-driven method for multiplicities $5$ to $10$ differ by up to $60\%$ from the simulated rates, as shown in Figure~\ref{fig:compare_radioactivity_rates}. These rates also significantly differ from the values used in the previous CCSN analysis~\cite{van2021km3net}, which were exclusively based on previous, simplified, simulation tools. Although these differences are expected to have a negligible impact on the results from~\cite{van2021km3net}, they could sizably affect multivariate analyses, such as the one presented in Sections~\ref{sec:filtering} and \ref{sec:sensitivity}, where this method is used to evaluate systematic uncertainties on the CCSN signal acceptance (i.e. the fraction of simulated CCSN events that survive the selection), and to estimate background levels for the final ARCA and ORCA detector configurations. Improving low-energy background estimates using single-DOM observables will allow for refining reconstruction algorithms and reducing systematic uncertainties across various KM3NeT searches.

\section{Single-DOM event selection for CCSN neutrino searches}
\label{sec:filtering}

In this section, a strategy is presented to identify the single-DOM signatures of CCSN neutrino interactions using Boosted Decision Trees.

\subsection{Identifying CCSN signals and selecting observables using Boosted Decision Trees}
\label{subsec:bdts}


Boosted Decision Trees (BDTs) are trained to separate CCSN neutrino signatures from the background events that survive the triggered event veto. The signal is simulated assuming the $11~M_\odot$ model of~\cite{tamborra2014neutrino}; the choice of supernova model has been found to have little impact on the performance. Data from the ORCA6 and ARCA21 detector configurations are used for modeling the background. Since the background composition in the multiplicity range $5-9$ strongly depends on both the detector structure and the multiplicity, five BDTs are used for ARCA and ORCA separately: one for each multiplicity from $5$ to $8$ and one for multiplicities $\geq 9$.

The BDTs employed for this analysis are built with the gradient boosting method of the XGBoost library~\cite{Chen:2016:XST:2939672.2939785}. For each multiplicity and detector configuration, the BDTs are trained using various subsets of the observables
introduced in Section~\ref{subsec:definitions} as features, and the classification accuracy is optimized using a mean-squared-error loss function. The predictability of the BDT is ensured through cross-validation, with a $3$:$1$ splitting between the training and test samples. Finally, the BDT hyperparameters are optimized by maximizing the background rejection for a benchmark signal acceptance of 50\%.

\begin{figure}[t]
    \centering
    \includegraphics[width=0.7\linewidth]{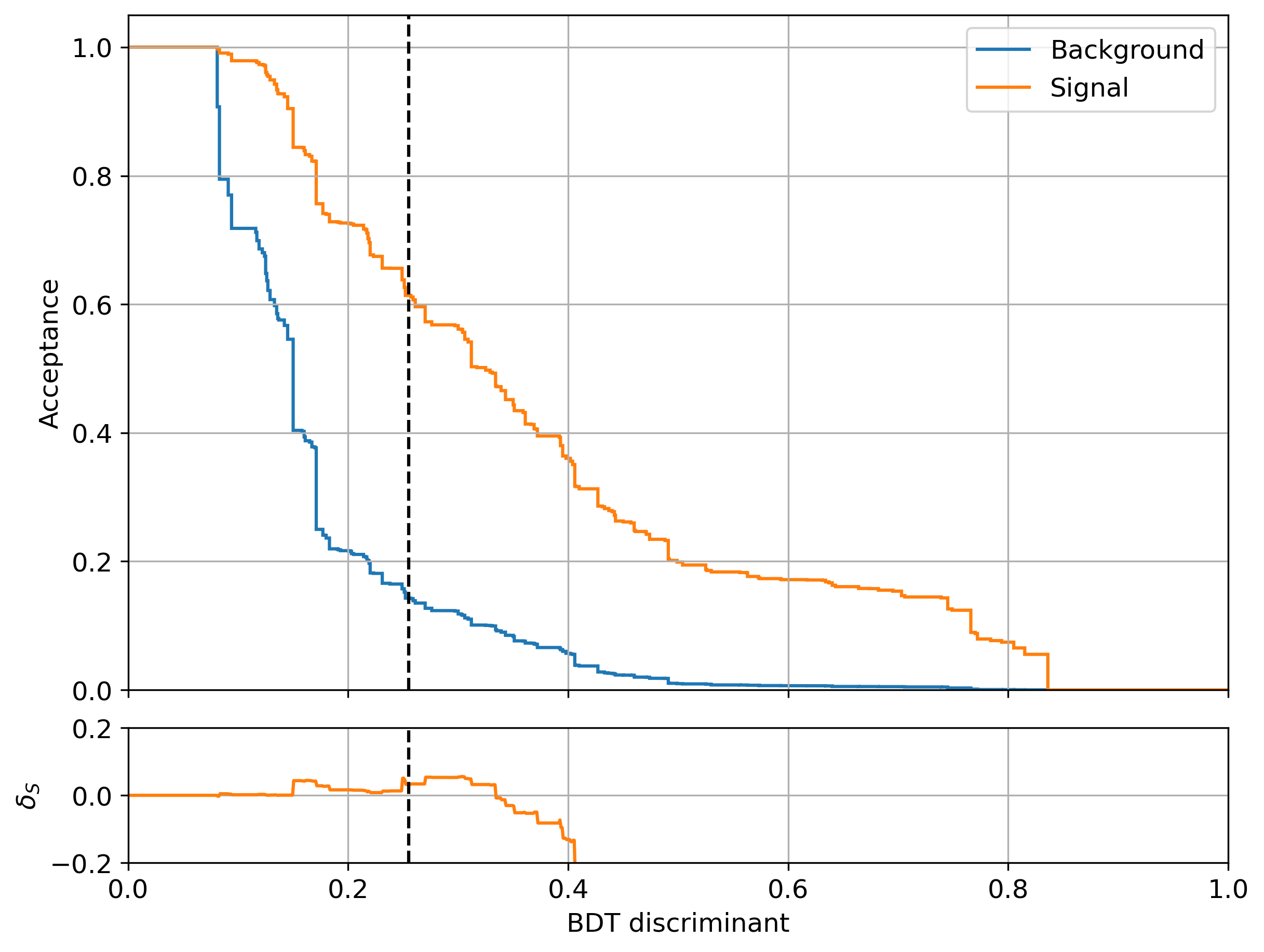}
    \caption{Acceptance of the BDT selection for ARCA21 events of multiplicity $7$, for simulated signal (orange) and background modelled using data (blue). The optimal cut derived in Section~\ref{subsec:cutopt} for this detector configuration is shown by the vertical line. The lower panel shows the relative systematic uncertainty on the BDT acceptance for the signal (obtained by comparing simulated backgrounds to data for a dedicated event sample as described in Section~\ref{subsec:systs}).
    }
    \label{fig:signal_syst}
\end{figure}

An example of the performance of the BDTs trained for this work is illustrated in Figure~\ref{fig:signal_syst}, for the ARCA21 detector. For all detector configurations and multiplicities, the use of a BDT significantly enhances KM3NeT's ability to identify CCSN neutrinos. Moreover, training BDTs on subsets of single-DOM observables demonstrated that the physics of MeV-scale signals can be effectively captured with only four variables: $|R|$, $\cos\theta$, total ToT, and $\langle\Delta t\rangle$. These four observables will hence be the only ones used for the rest of the analysis.

\subsection{Systematic uncertainties}
\label{subsec:systs}

Inaccuracies in the modeling of signal and background single-DOM signatures will lead to systematic uncertainties on the signal and background acceptances. Both are evaluated by comparing simulated background events to data, taking the associated relative difference as the measure of the systematic uncertainty. The simulations are corrected using the data-driven procedure described in Section~\ref{sec:km3bg}. 

\begin{figure}[t]
    \centering
    \includegraphics[width=0.495\linewidth]{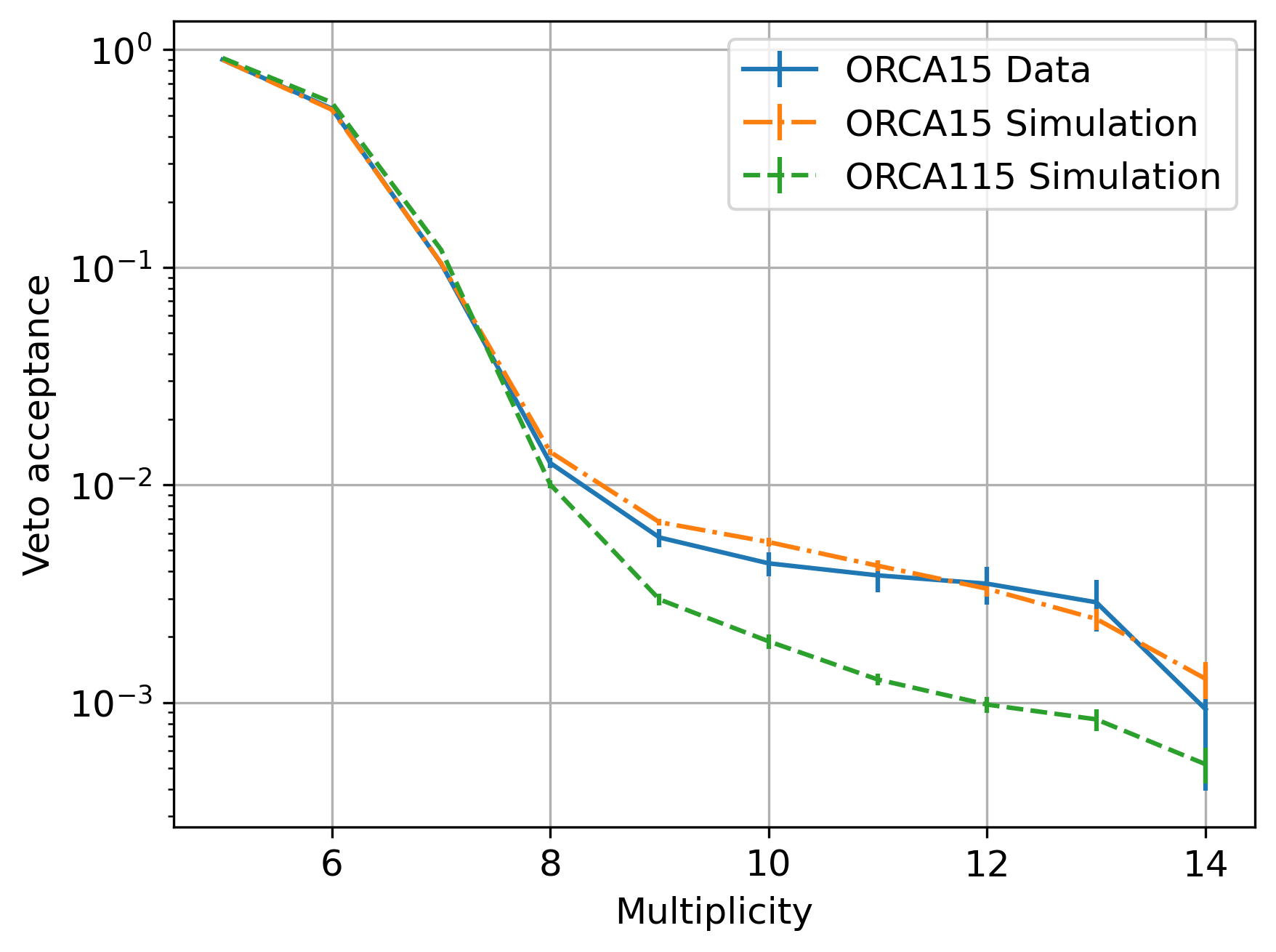}  \includegraphics[width=0.487\linewidth]{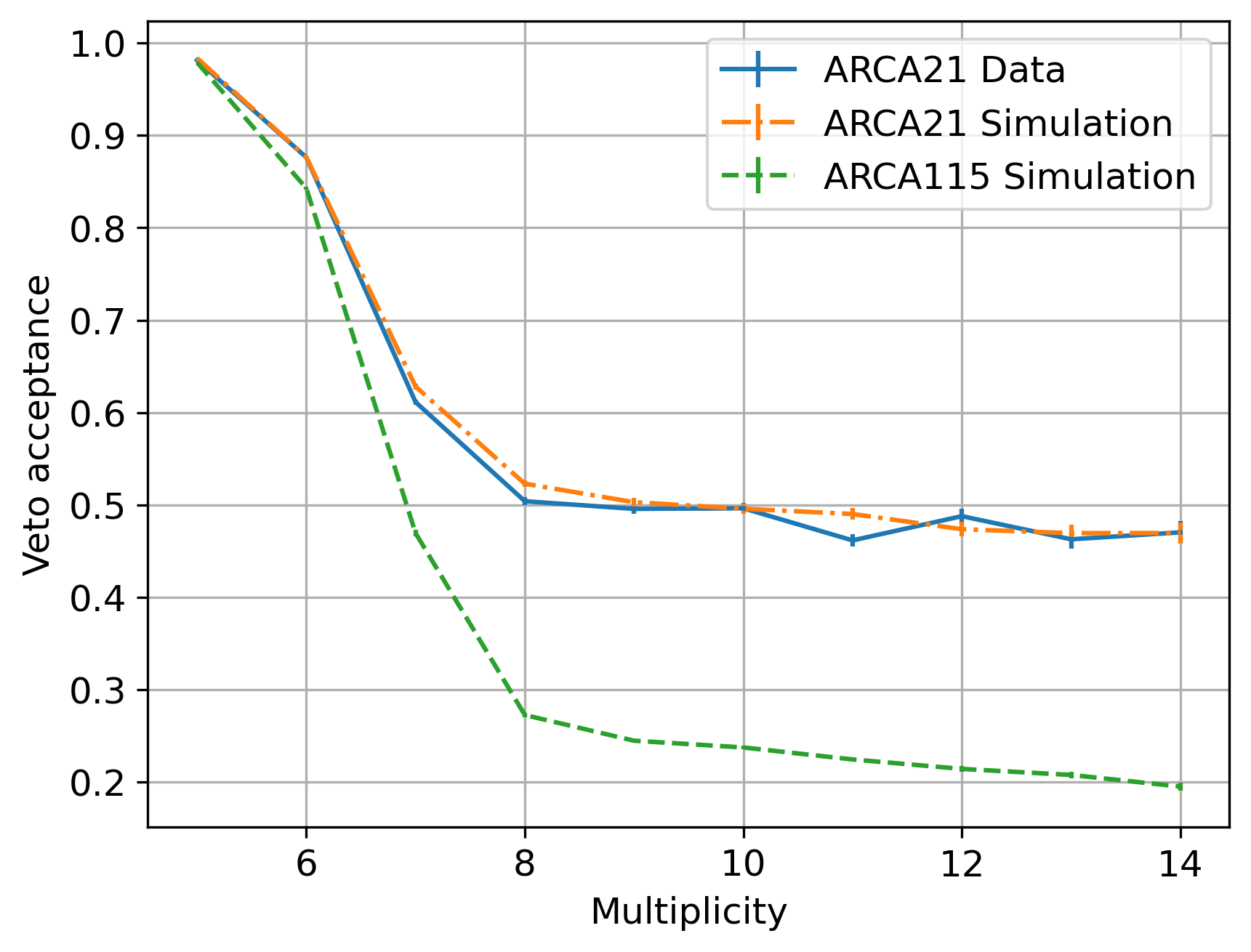}
    \caption{Veto acceptances for ORCA (left) and ARCA (right). For partial ARCA and ORCA configurations, the data and the atmospheric muon simulations are shown as solid blue and dash-dotted orange lines, respectively. The predicted veto acceptances for ORCA115 and one 115-DU building block of ARCA230 are shown as dashed green lines.}
\label{fig:muon_veto_acceptances}
\end{figure}

The main source of uncertainties on the background rates is the extrapolation of the fraction of background events surviving the triggered event veto, or ``veto acceptance'', when considering future detector configurations. Radioactivity backgrounds are not affected by the triggered event veto, whereas atmospheric muon backgrounds are strongly suppressed in a manner that is dependent on the detector geometry~\cite{van2021km3net}. Consequently, for a given event sample, the veto acceptance is determined by two quantities: (1) the relative fraction of radioactivity events in the sample before the veto, which is obtained through the data-driven method described in Section~\ref{sec:km3bg}, and (2) the fraction of atmospheric muon events that are found to survive the veto, evaluated using simulations. Following the procedure from~\cite{van2021km3net}, uncertainties on the veto acceptance are estimated by
 comparing data and simulations for the ORCA15 and ARCA21 configurations for each value of the multiplicity and the BDT discriminant. To avoid underestimating the systematic uncertainties, simulations are corrected before the triggered event veto is applied.
The veto acceptance as a function of the multiplicity is presented in Figure~\ref{fig:muon_veto_acceptances} for both ORCA and ARCA, showing that discrepancies between data and simulations and the resulting systematic uncertainties are well under control.
 
To evaluate systematic uncertainties on the CCSN signal acceptance, background simulations (see Appendix~\ref{subsec:SignalSimu}) are compared to data. Since the properties of the signal events do not depend on the detector configuration, data and simulations are compared for a single ORCA6 run. As demonstrated in Figure~\ref{fig:signal_syst}, systematic uncertainties remain within $10\%$ for low and intermediate cut values, but increase for high values of the BDT discriminant. This behavior is therefore taken into account during the cut optimization procedure.

Systematic uncertainties on the neutrino interaction cross section, water optical properties, and PMT characteristics are also considered both for the cut optimization and when computing sensitivities to CCSN models. The uncertainty values from~\cite{van2021km3net} are used for the cut optimization, and these values are re-evaluated when computing the sensitivities.

\section{KM3NeT sensitivity to CCSN neutrinos} 
\label{sec:sensitivity}

A key objective of this analysis is to position KM3NeT as a strong contributor to the Supernova Early Warning System (SNEWS)~\cite{al2021snews}, which combines the information received from multiple detectors to reliably identify a CCSN (with a false alarm rate of one per century), locate the supernova using triangulation, and notify telescopes for prompt electromagnetic follow-up. The procedure developed in this paper is designed to identify potential CCSN neutrino signals and issue rapid alerts to this global network of observatories and detectors, following the realtime analysis procedure described in~\cite{aiello2022implementation}. In this section, the sensitivity of the new CCSN analysis based on BDTs, incorporating the selection steps and systematic uncertainty evaluation procedure described above, is estimated for different detector configurations and CCSN models. The number of selected single-DOM events is computed within a \SI{0.5}{s} time window, corresponding to the high-rate neutrino emission phase predicted by most models, and compared to the background expectation. As in the previous analysis~\cite{van2021km3net,aiello2022implementation}, the selection steps are optimized to maximize the distance up to which a CCSN with a neutrino flux corresponding to the $11~M_\odot$ model from~\cite{tamborra2014neutrino} can be detected at $5\sigma$.

\subsection{Selection procedure}
\label{subsec:cutopt}

In the previous CCSN analysis~\cite{van2021km3net}, the optimal multiplicity range was defined using a simple analytical approximation of the signal detection significance, neglecting systematic uncertainties. In this analysis, the signal detection significance is evaluated using the Rolke method~\cite{rolke2005limits}, which accounts for systematic uncertainties.  

Four detector configurations have been considered to optimize the BDT and multiplicity cuts for a CCSN spectrum corresponding to the $11~M_\odot$ model considered in~\cite{van2021km3net}: ARCA29, ORCA24, ARCA230 and ORCA115. The former represent the first ARCA and ORCA configurations for which the method presented here was used to search for CCSNe in real time. The latter represent the final ARCA and ORCA configurations. In each case, the background is modeled using data from the ARCA21 and ORCA15 configurations, respectively, and the signal is simulated using the average PMT efficiencies associated with these detectors. Since the typical $f_A$ values for ORCA15 range between $97$ and $98\%$, the associated background rates have been rescaled to model the $f_A > 99\%$ scenario. With this choice of detector configurations, for ARCA29 and ORCA24, background rates can be rescaled linearly to account for the increase in detector size, as the variation of the background rate per DOM between the different configurations is negligible. 
For ARCA230 and ORCA115, an additional step is required to account for the reduced veto acceptance for background muons resulting from the increase in detector size. For this, the background evaluation method described in~\cite{van2021km3net} is used: the background rates are evaluated from data before applying the triggered event veto, rescaled to match the increase in the number of DOMs, and then multiplied by the veto acceptances obtained from simulations (shown in Figure~\ref{fig:muon_veto_acceptances}).  

\begin{figure}
    \centering
    \includegraphics[width=0.9\linewidth]{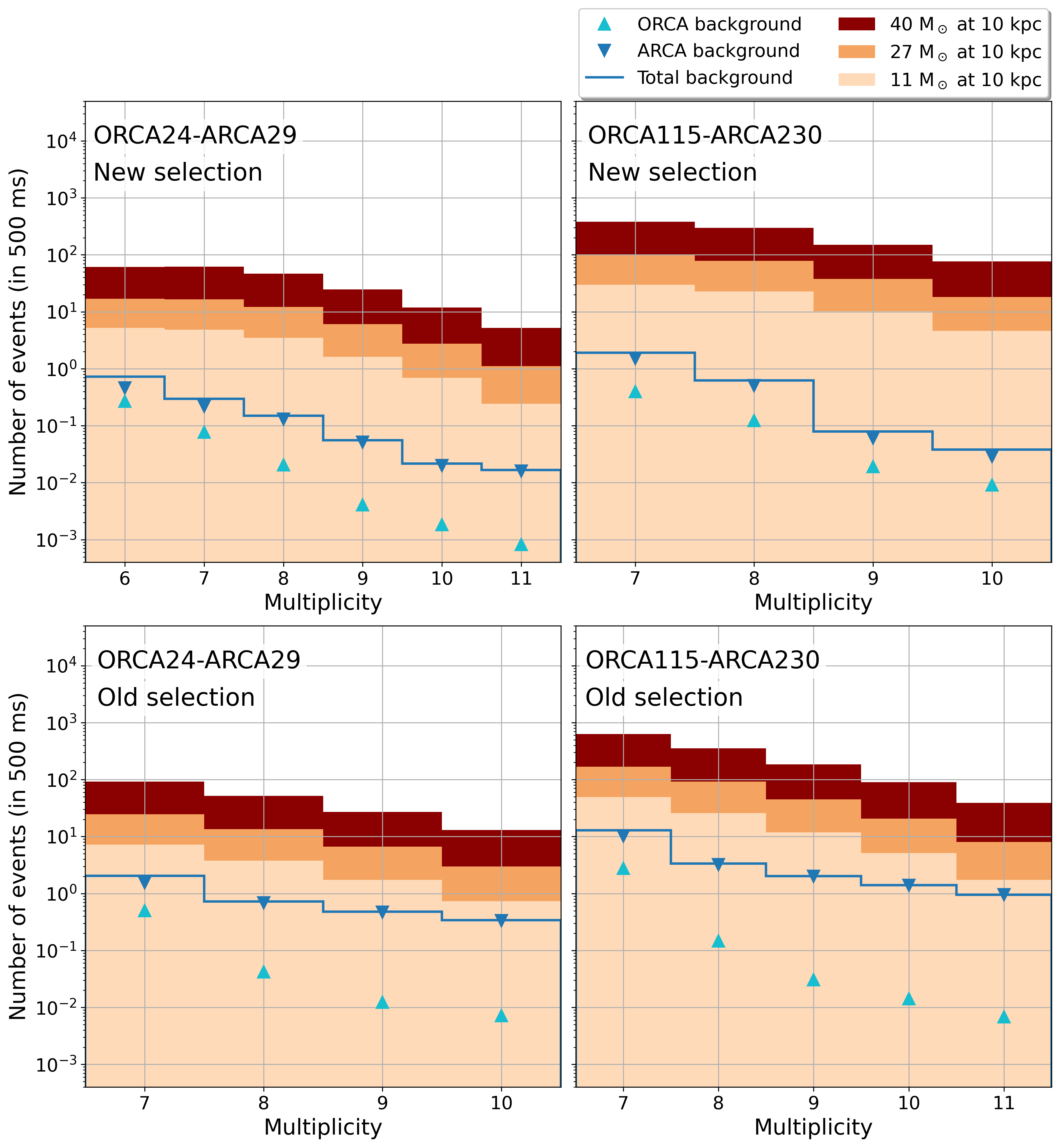}
    \caption{Numbers of selected CCSN candidates for signal and background using the BDT-based method presented in this article (top) and the method from~\cite{van2021km3net} (bottom), as a function of the multiplicity for ORCA24-ARCA29 (left) and ORCA115-ARCA230 (right). The multiplicity range shown for each detector configuration and selection procedure is the one providing the optimal sensitivity, as discussed in Section~\ref{subsec:cutopt}. For the signal, the $11~M_\odot$, $27~M_\odot$, and $40~M_\odot$ models used in~\cite{van2021km3net} are shown. The light-blue upward-pointing and dark-blue downward-pointing triangles represent the ORCA and ARCA backgrounds, respectively. The dark blue histogram is the sum of the two backgrounds.}
    \label{fig:final_multiplicities}
\end{figure}

The numbers of selected CCSN candidates after the optimization procedure are shown in the top panel of Figure~\ref{fig:final_multiplicities} for the ORCA24 and ARCA29 detector configurations and for ORCA115-ARCA230. The optimal multiplicity ranges for the ORCA24-ARCA29 and ORCA115-ARCA230 detector configurations are $6-11$ and $7-10$, respectively, close to the range found in~\cite{van2021km3net}. As in the previous KM3NeT CCSN analysis, ORCA backgrounds are less dominant thanks to ORCA's higher atmospheric muon identification capabilities at low energies. For the signal, the $11~M_\odot$, $27~M_\odot$, and $40~M_\odot$ models used in~\cite{van2021km3net} are shown for a CCSN at \SI{10}{\kilo\parsec}. For all three models, the signal is dominating over the background, indicating that even with a partial configuration, KM3NeT is sensitive to most CCSN candidates in the dense Galactic bulge. To assess the impact of the BDT selection, the signal and background distributions obtained before this selection are shown in the bottom panel of Figure~\ref{fig:final_multiplicities}. The background rates differ from~\cite{van2021km3net} as they have been evaluated using the method described in Section~\ref{sec:bkgmodels} but the resulting impact on KM3NeT's detection horizon is negligible. Applying BDT selection criteria significantly improves the signal-to-background ratio, in particular by improving the muon rejection rate at ARCA. 

\subsection{Sensitivities to CCSN models}
\label{subsec:sensitivity}

 The number of detected single-DOM signal events associated with a given significance is estimated using the same Rolke method as in Section~\ref{sec:sensitivity}. The systematic uncertainties associated with the selection cuts are summarized in Table~\ref{tab:uncertainties}.

\begin{table}[h!]
\begin{center}
\begin{threeparttable}
\caption{Systematic uncertainties of the analysis. The third column indicates the resulting systematic uncertainties on the signal and background rates denoted as (S) and (B), respectively. The evaluation of the uncertainties on the muon veto acceptance and single-DOM signal event modeling is described in Section~\ref{subsec:systs}. These uncertainties are quoted here for the optimal BDT cuts corresponding to the recent and final detector configurations. The monthly variability of the instrumentation efficiency~\cite{aiello2022implementation} has been evaluated over a period of $1$ year.  The variations and uncertainties on the water transparency, the finite generation volume and the inverse beta decay (IBD) and elastic scattering (ES) cross sections are taken from~\cite{van2021km3net}. The impact of the cross-section uncertainties for neutrino charged-current interactions with $^{16}$O are neglected, as these processes account for less than $1\%$ of the detected events.}
\label{tab:uncertainties}
\begin{tabular}{cccc}
\hline
\textbf{Source of the uncertainty} & \textbf{Variation} & \multicolumn{2}{c}{\textbf{Systematic uncertainty}} \\
\hline
\multicolumn{2}{c}{ } & ORCA24-ARCA29 & ORCA115-ARCA230\\
\cline{3-4}
Muon veto acceptance &  & (B) N/A & $6.0$\%\\
Single-DOM signal event modeling &  & (S) $7.9\%$ & $8.4$\%\\
Instrumentation efficiency variability$^1$
&  $1$ year & \multicolumn{2}{c}{(S,B) $11\%$} \\
PMT efficiency & $\pm 5 \%$ & \multicolumn{2}{c}{(S) $11 \%$}\\
Water transparency & $\pm 10 \%$ & \multicolumn{2}{c}{(S) $1 \%$} \\
Finite generation volume & $+\SI{5}{\m}$ & \multicolumn{2}{c}{(S) $<1\%$} \\
IBD/ES cross sections & $<1\%$ & \multicolumn{2}{c}{(S) $<1\%$}\\
\hline
\end{tabular}
\begin{tablenotes}
\item[1] \small{In~\cite{van2021km3net}, the source of the corresponding uncertainty was considered to be the fraction of active PMTs. ``Instrumentation efficiency variability'' is a more accurate terminology which was introduced in a subsequent study~\cite{aiello2022implementation}.}
\end{tablenotes}
\end{threeparttable}
\end{center}
\end{table}

\begin{figure}
    \centering
    \includegraphics[width=0.49\linewidth]{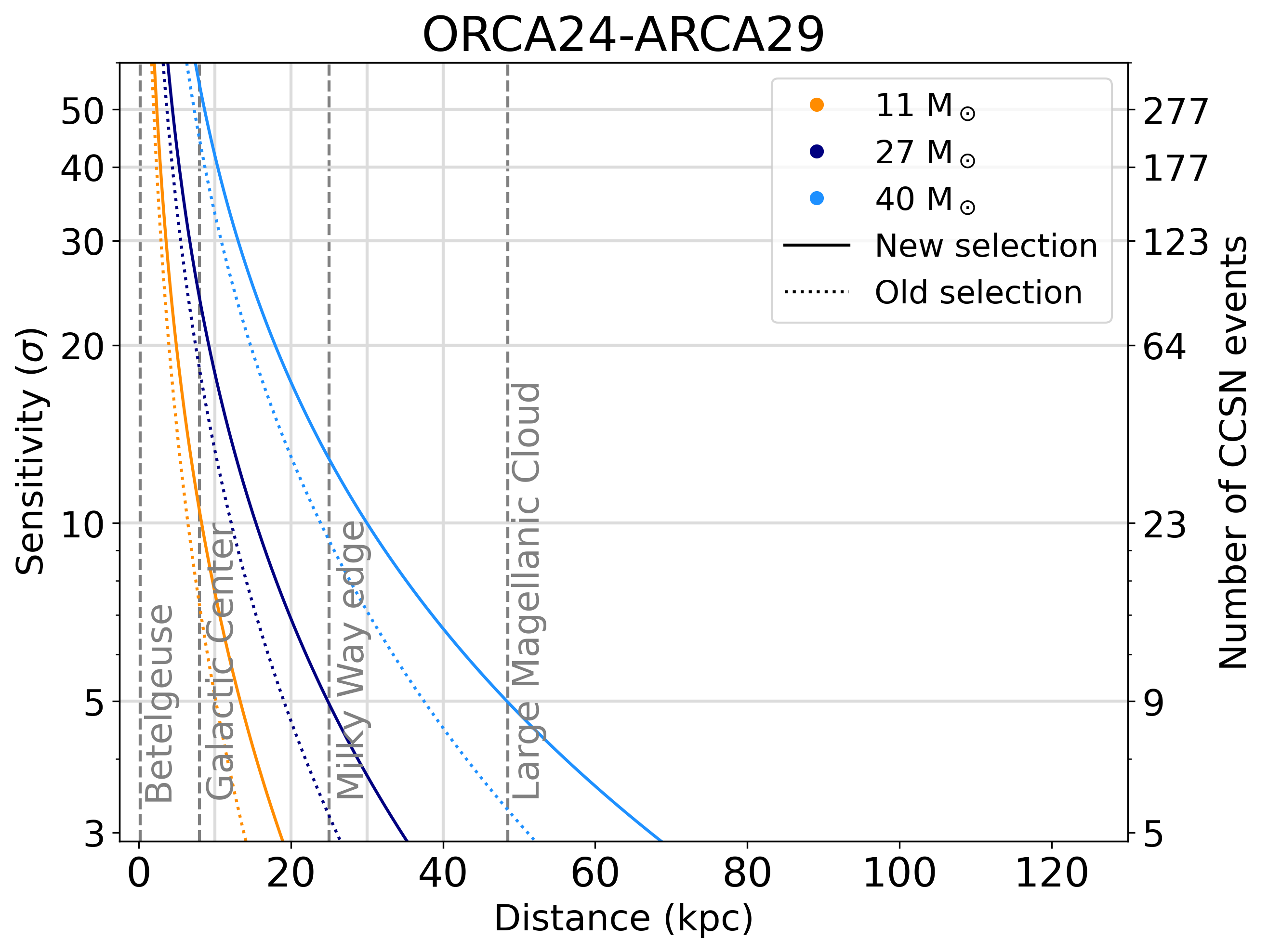}
    \includegraphics[width=0.49\linewidth]{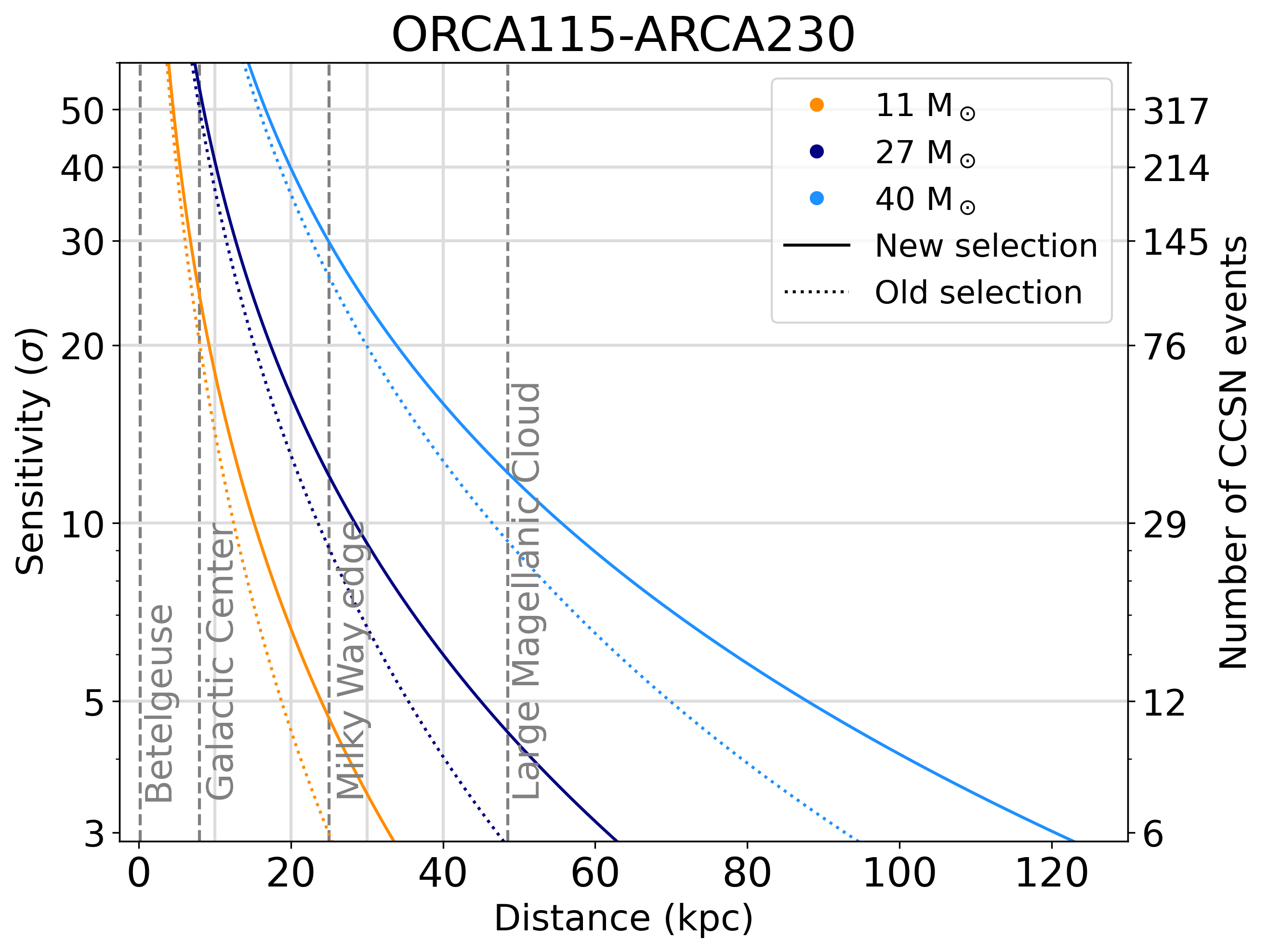}
    \caption{Sensitivity, in units of $\sigma$, as a function of the CCSN distance to Earth for the $11~M_\odot$~\cite{tamborra2014neutrino}, $27~M_\odot$~\cite{tamborra2014neutrino}, and $40~M_\odot$~\cite{walk2020neutrino} models already considered in the previous KM3NeT CCSN search~\cite{van2021km3net}. Left: sensitivities for the ORCA24-ARCA29 detector configuration. Right: projected sensitivities for the completed ORCA115-ARCA230 detector. The sensitivities obtained with the selection from ~\cite{van2021km3net} are indicated by the dotted lines. For ORCA115-ARCA230, these detection horizons agree with the ones from~\cite{van2021km3net} despite the different background estimation method.}
    \label{fig:sensitivity}
\end{figure}

Sensitivities are evaluated for the CCSN signals in the case of the $11~M_\odot$, $27~M_\odot$, and $40~M_\odot$ progenitor mass models and are shown in Figure~\ref{fig:sensitivity}, as a function of the distance to the CCSN, for the ORCA24-ARCA29 (left) and ORCA115-ARCA230 (right) detector configurations. Introducing the BDT selection leads to a $25$ to $30\%$ increase of the $5\sigma$ distance horizon as compared to the one obtained with the method from~\cite{van2021km3net}. For the $11~M_\odot$ CCSN model, associated with the lowest neutrino flux, KM3NeT's recent partial configurations have been sensitive to CCSNe up to a distance of \SI{13}{\kilo\parsec} and the detector will achieve full Galactic sensitivity when completed. The current reach of KM3NeT allows probing at least $79\%$ of Galactic CCSN candidates~\cite{Adams:2013ana}, $46\%$ more than with the previous analysis strategy~\cite{van2021km3net}. The results highlight both KM3NeT's potential for characterizing MeV neutrino bursts and the interest of investigating the structure of single-DOM hit coincidences to study low-energy signals. 

\begin{figure}
    \centering
    \includegraphics[width=0.49\linewidth]{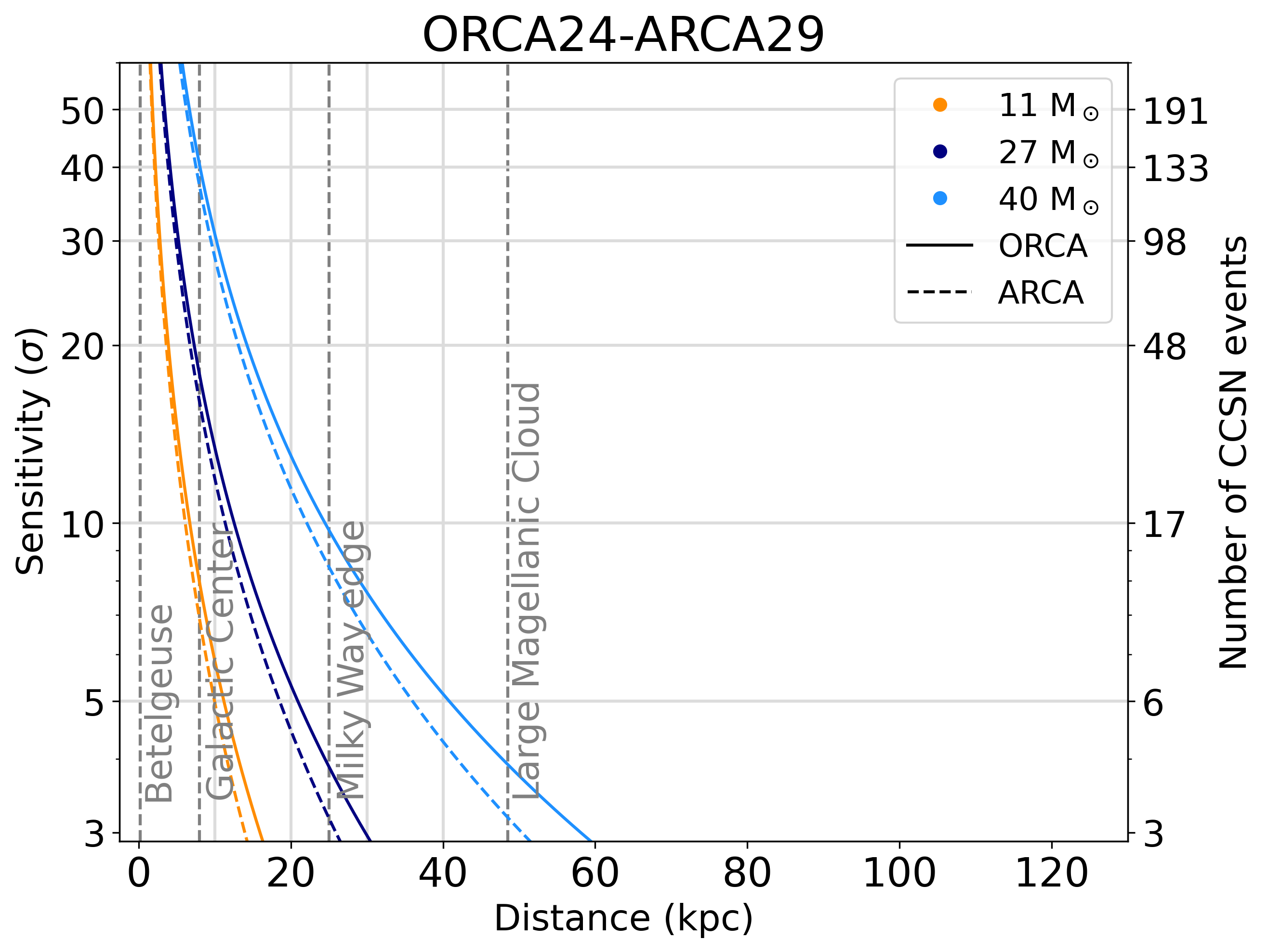}
    \includegraphics[width=0.49\linewidth]{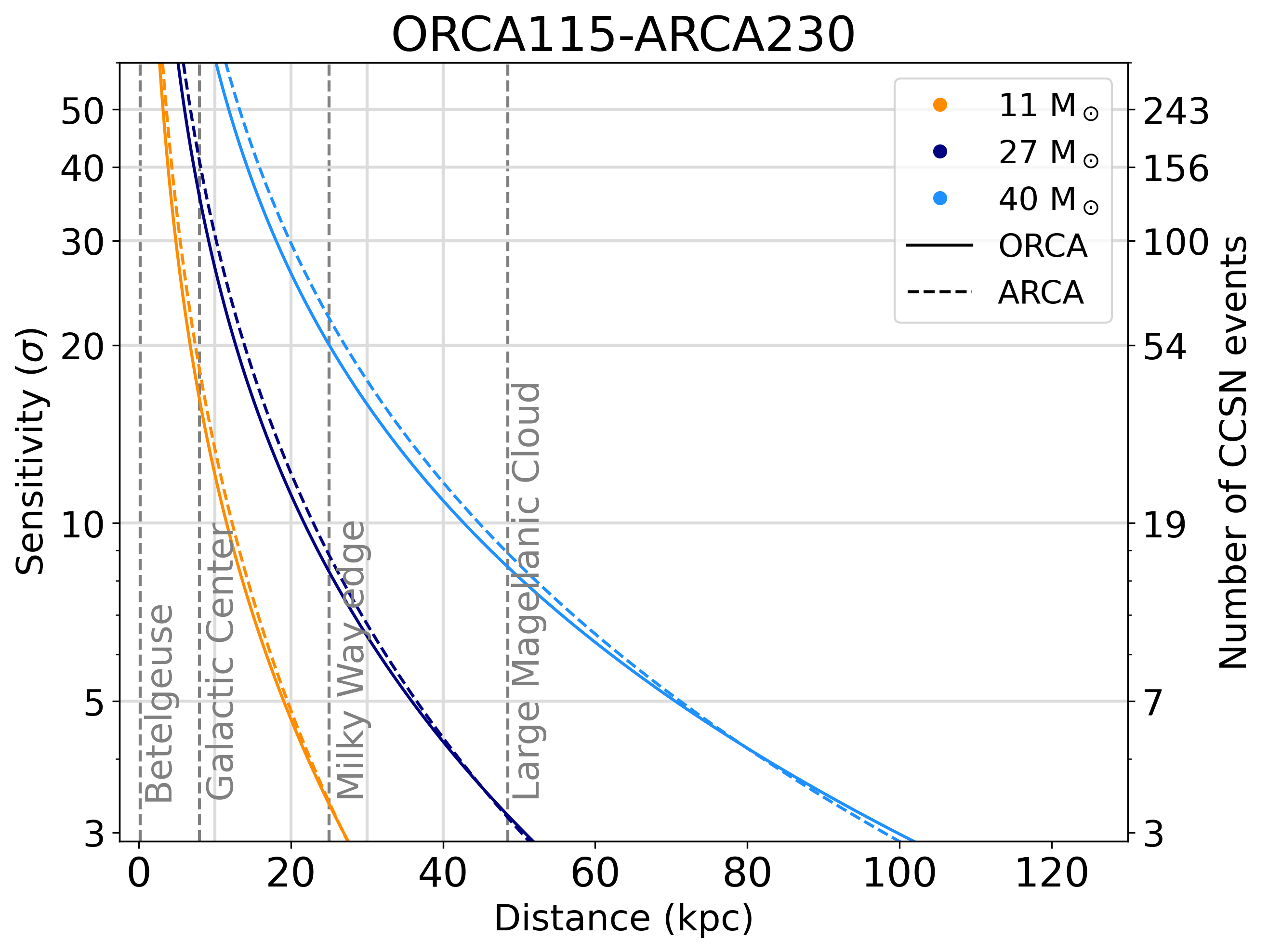}
    \caption{Sensitivity, in units of $\sigma$, as a function of the CCSN distance to Earth for the $11~M_\odot$~\cite{tamborra2014neutrino}, $27~M_\odot$~\cite{tamborra2014neutrino}, and $40~M_\odot$~\cite{walk2020neutrino} models already considered in the previous KM3NeT CCSN search~\cite{van2021km3net}. In this figure, the performances of ARCA and ORCA are evaluated separately. The sensitivities for the ORCA24 (solid lines) and ARCA29 (dashed lines) detector configurations are indicated in the left panel, and the projected sensitivities for the ORCA115 (solid lines) and ARCA230 (dashed lines) detectors are indicated in the right panel.}
    \label{fig:sensitivity_ORCAandARCA}
\end{figure}

The $5\sigma$ CCSN detection horizons for the considered ARCA and ORCA configurations have also been evaluated for each detector separately and are shown in Figure~\ref{fig:sensitivity_ORCAandARCA}. For the $11~M_\odot$ CCSN model, the $5\sigma$ detection horizons for ORCA24 and ORCA115 are \SI{11.2}{\kilo\parsec} and \SI{19.0}{\kilo\parsec}, respectively. The horizons for ARCA29 and ARCA230 are similar: \SI{10.0}{\kilo\parsec} and \SI{19.6}{\kilo\parsec}, respectively. Despite the different numbers of DOMs in the two detectors, their performances are comparable because the larger size of ARCA is compensated by the superior performance of the triggered event veto at ORCA. Thus, ORCA is the best-performing detector when the number of expected CCSN neutrino interactions is low, e.g. for the partial detector configurations, while ARCA is more efficient for close-by supernovae, for which the background is negligible. Overall, each detector is currently sensitive to at least $57\%$ of supernova candidates in the Milky Way, and will reach nearly full Galactic sensitivity once completed. Since ARCA and ORCA work independently, these results show that KM3NeT will have a high duty cycle for supernova detection. 

\subsection{Impact of variations in data-taking conditions}
\label{sec:DetectorConditions}

\begin{figure}
    \centering
    \includegraphics[width=0.7\linewidth]{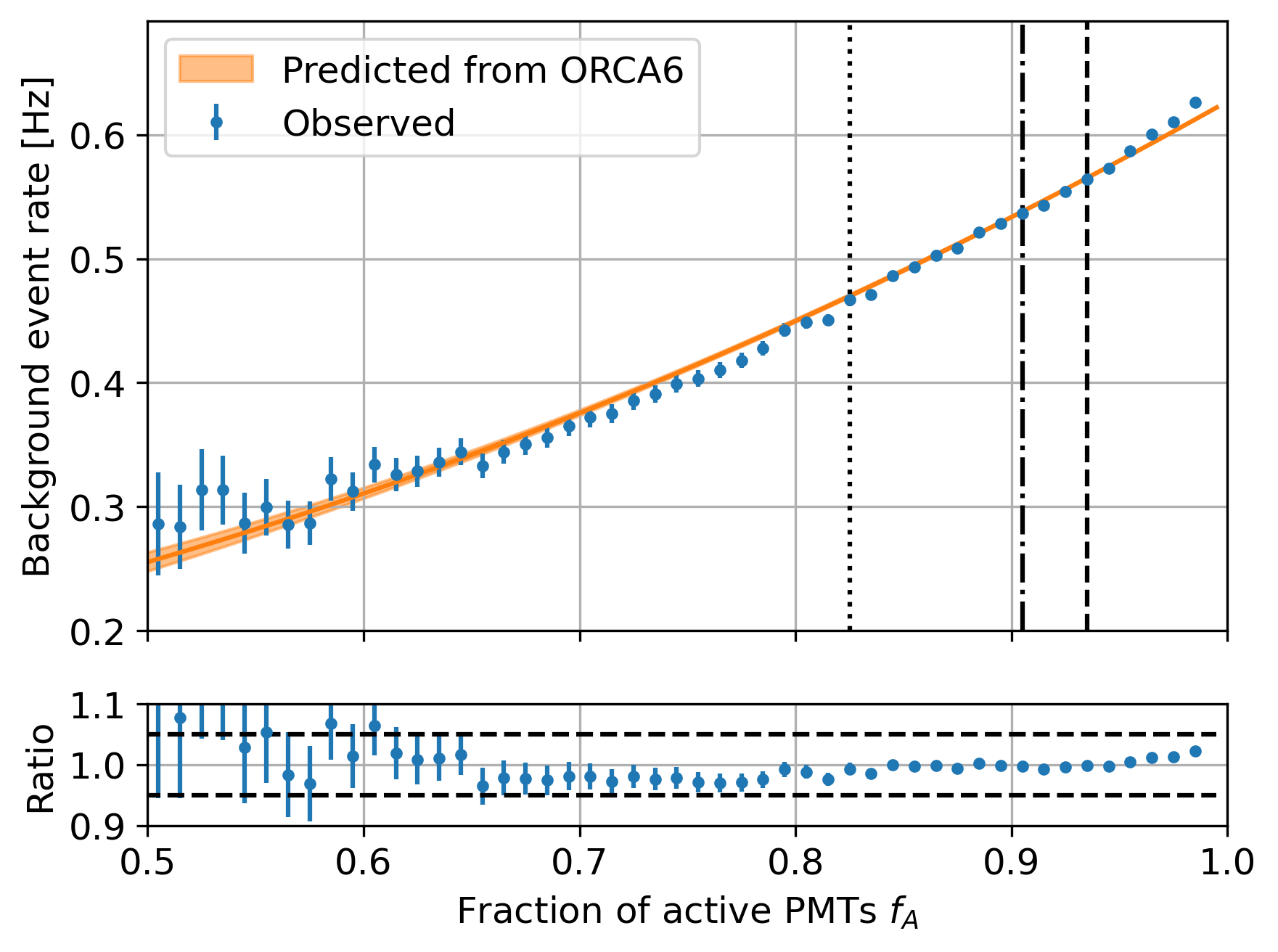}
    \caption{Background rates for the CCSN analysis as a function of the fraction of active PMTs $f_A$. The blue dots with error bars indicate the rates measured at ORCA18 and the orange band shows the rates predicted using a parameterization based on ORCA6 data. The regions to the right of the dashed, dash-dotted, and dotted lines correspond to $50\%$, $68\%$, and $90\%$ of the data-taking time, respectively. The ratio of the observed and predicted rates is represented in the lower panel, with the dashed lines indicating $\pm 5\%$ differences.}
    \label{fig:rate_fit_hrv}
\end{figure}

The sensitivities presented above have been evaluated assuming a fraction of active PMTs $f_A > 99\%$. While this value is now achieved for more than $99\%$ of ARCA's data-taking period, $f_A$ can reach values as low as $50\%$ at ORCA, where bioluminescence levels are typically higher. To account for these variations, the parameterization of the background rate dependence on $f_A$ proposed in \cite{aiello2022implementation} is adapted to the BDT-based single-DOM event selection procedure. 
The rate of single-DOM background events as a function of $f_A$ is shown in Figure~\ref{fig:rate_fit_hrv} for ORCA18. For more than $95\%$ of ORCA18's data-taking period, background rates can be successfully predicted using a parameterization obtained for the ORCA6 configuration. This predictability enables a reliable estimation of the expected background level when searching for CCSNe in real time, even when the bioluminescence activity is high. For ORCA, during $68\%$ of the data-taking time, $f_A$ will range from $91\%$ to $99\%$, resulting in variation of ORCA's CCSN detection horizon by only $8\%$. The variation of the combined ORCA+ARCA detection horizon is expected to be even smaller due to the lower bioluminescence activity (hence typically larger $f_A$) at ARCA. Hence, KM3NeT's CCSN identification capabilities are robust even in the highly dynamic environment of the Mediterranean Sea.   

\section{Conclusions}
\label{sec:conclusion}
In this paper, a novel approach is proposed to improve the sensitivity of KM3NeT to CCSN neutrinos, by selecting single-DOM hit clusters based on their topology. A multi-variate analysis using BDTs has demonstrated that most of the hit information in a single-DOM event can be encapsulated into a small set of observables. With these observables, the ability to  identify CCSN neutrino signals and separate them from radioactivity and atmospheric muon signatures, thereby allowing the estimation of the corresponding systematic uncertainties, is enhanced. Consequently, KM3NeT's CCSN detection horizon is extended by $30\%$ for both current and future detector configurations and variations in the detector’s performance due to bioluminescence activity remain below $10$\% for most of the data-taking time. When implemented in KM3NeT's realtime analysis system, the new, BDT-based, method led to a $46$\% increase of the detection horizon. Even for the most pessimistic model considered here, the current detector configurations could probe the vast majority of CCSN candidates in the Milky Way and, once completed, KM3NeT will achieve full Galactic sensitivity.  Hence, even with its current partial configuration, KM3NeT can play an important role in the detection and localization of the next Galactic supernova, in particular as part of the Supernova Early Warning System~\cite{al2021snews}. The results of this study highlight the versatile performance of the multi-PMT design to characterize environmental backgrounds and investigate low-energy phenomena at neutrino telescopes deployed in natural, dynamic environments.

\section{Acknowledgements}
The authors acknowledge the financial support of:
KM3NeT-INFRADEV2 project, funded by the European Union Horizon Europe Research and Innovation Programme under grant agreement No 101079679;
Funds for Scientific Research (FRS-FNRS), Francqui foundation, BAEF foundation.
Czech Science Foundation (GAČR 24-12702S);
Agence Nationale de la Recherche (contract ANR-15-CE31-0020), Centre National de la Recherche Scientifique (CNRS), Commission Europ\'eenne (FEDER fund and Marie Curie Program), LabEx UnivEarthS (ANR-10-LABX-0023 and ANR-18-IDEX-0001), Paris \^Ile-de-France Region, Normandy Region (Alpha, Blue-waves and Neptune), France,
The Provence-Alpes-Côte d'Azur Delegation for Research and Innovation (DRARI), the Provence-Alpes-Côte d'Azur region, the Bouches-du-Rhône Departmental Council, the Metropolis of Aix-Marseille Provence and the City of Marseille through the CPER 2021-2027 NEUMED project,
The CNRS Institut National de Physique Nucléaire et de Physique des Particules (IN2P3);
Shota Rustaveli National Science Foundation of Georgia (SRNSFG, FR-22-13708), Georgia;
This research was funded by the European Union (ERC MuSES project No 101142396); 
The General Secretariat of Research and Innovation (GSRI), Greece;
Istituto Nazionale di Fisica Nucleare (INFN) and Ministero dell’Universit{\`a} e della Ricerca (MUR), through PRIN 2022 program (Grant PANTHEON 2022E2J4RK, Next Generation EU) and PON R\&I program (Avviso n. 424 del 28 febbraio 2018, Progetto PACK-PIR01 00021), Italy; IDMAR project Po-Fesr Sicilian Region az. 1.5.1; A. De Benedittis, W. Idrissi Ibnsalih, M. Bendahman, A. Nayerhoda, G. Papalashvili, I. C. Rea, A. Simonelli have been supported by the Italian Ministero dell'Universit{\`a} e della Ricerca (MUR), Progetto CIR01 00021 (Avviso n. 2595 del 24 dicembre 2019); KM3NeT4RR MUR Project National Recovery and Resilience Plan (NRRP), Mission 4 Component 2 Investment 3.1, Funded by the European Union – NextGenerationEU,CUP I57G21000040001, Concession Decree MUR No. n. Prot. 123 del 21/06/2022;
Ministry of Higher Education, Scientific Research and Innovation, Morocco, and the Arab Fund for Economic and Social Development, Kuwait;
Nederlandse organisatie voor Wetenschappelijk Onderzoek (NWO), the Netherlands;
The grant “AstroCeNT: Particle Astrophysics Science and Technology Centre”, carried out within the International Research Agendas programme of the Foundation for Polish Science financed by the European Union under the European Regional Development Fund; The program: “Excellence initiative-research university” for the AGH University in Krakow; The ARTIQ project: UMO-2021/01/2/ST6/00004 and ARTIQ/0004/2021;
Ministry of Research, Innovation and Digitalisation, Romania;
Slovak Research and Development Agency under Contract No. APVV-22-0413; Ministry of Education, Research, Development and Youth of the Slovak Republic;
MCIN for PID2021-124591NB-C41, -C42, -C43 and PDC2023-145913-I00 funded by MCIN/AEI/10.13039/501100011033 and by “ERDF A way of making Europe”, for ASFAE/2022/014 and ASFAE/2022 /023 with funding from the EU NextGenerationEU (PRTR-C17.I01) and Generalitat Valenciana, for Grant AST22\_6.2 with funding from Consejer\'{\i}a de Universidad, Investigaci\'on e Innovaci\'on and Gobierno de Espa\~na and European Union - NextGenerationEU, for CSIC-INFRA23013 and for CNS2023-144099, Generalitat Valenciana for CIDEGENT/2020/049, CIDEGENT/2021/23, CIDEIG/2023/20, ESGENT2024/24, CIPROM/2023/51, GRISOLIAP/2021/192 and INNVA1/2024/110 (IVACE+i), Spain;
Khalifa University internal grants (ESIG-2023-008, RIG-2023-070 and RIG-2024-047), United Arab Emirates;
The European Union's Horizon 2020 Research and Innovation Programme (ChETEC-INFRA - Project no. 101008324).

Views and opinions expressed are those of the author(s) only and do not necessarily reflect those of the European Union or the European Research Council. Neither the European Union nor the granting authority can be held responsible for them. 

\newpage
\appendix
\section{Data samples for low-energy searches}
\label{subsubsec:samples}


To evaluate the sensitivity of KM3NeT to low-energy neutrinos, data samples corresponding to different detector geometries and specific $f_A$ ranges are built. In this study, only data-taking runs classified as ``good'', i.e. without major data-acquisition issues, are considered. For the background estimates and CCSN sensitivity studies presented in Sections~\ref{sec:CCSNtriggers}, \ref{sec:km3bg}, \ref{sec:filtering}, and \ref{sec:sensitivity}, the following three samples are used:
\begin{itemize}
    \item \texttt{ORCA6}: this is an ORCA configuration for which $f_A>99\%$ for a sizable fraction of the data-taking time. The sample is used to evaluate radioactivity and muon background rates for ORCA in Section~\ref{sec:km3bg} and to compute the systematic uncertainties on the CCSN signal efficiency in Section~\ref{sec:filtering}.
    \item \texttt{ARCA21}: for this configuration, the fraction of active PMTs is larger than $99\%$ for most of the data-taking time. The high-$f_A$ sample is used to evaluate radioactivity and muon background rates for ARCA in Section~\ref{sec:km3bg}, compute the systematic uncertainties on the CCSN signal efficiency and the muon veto acceptance in Section~\ref{sec:filtering}, and model the total background for CCSN sensitivity estimates in Section~\ref{sec:sensitivity}.
    \item \texttt{ORCA15}: here, SN timeslices with $97\% < f_A < 98\%$ are selected. This relaxed requirement on $f_A$ is due to the lack of observation periods with $f_A > 98\%$ for the recent ORCA configurations. The sample is used to evaluate the systematic uncertainties on the muon veto acceptance in Section~\ref{sec:filtering} and model the total background for CCSN sensitivity estimates in Section~\ref{sec:sensitivity}.
\end{itemize}
\section{Signal and background simulations}
\subsection{Signal simulation}
\label{subsec:SignalSimu}

The CCSN neutrino simulation for KM3NeT has been designed to enable the study of a wide range of supernovae, with different locations and neutrino energy spectra. For this, positrons are simulated with directions and energies uniformly distributed within $4\pi$ sr and $0$--\SI{100}{\mega\eV}, respectively. The detector modeling is reduced to that of a single DOM, since CCSN neutrino signatures spanning multiple DOMs are expected to be extremely rare~\cite{van2021km3net}. The production vertices of IBD positrons are taken to be uniformly distributed in space within a \SI{20}{\m} radius around the DOM. The corresponding volume has been shown in~\cite{van2021km3net} to contain more than $99\%$ of the single-DOM events with multiplicities greater than $1$. Light propagation, PMT response, and the generation of associated timeslices are then simulated using \texttt{GEANT4}-based KM3NeT simulation tools~\cite{tsirigotis2011hou, colomer2017detailed}. 

For a given supernova model, neutrino interaction process, and CCSN position in the sky, the corresponding supernova signal is modeled by weighting the simulated single-DOM events\footnote{Since electrons and positrons are indistinguishable in KM3NeT, the simulated positrons can be used to model electrons from e.g. neutrino-electron elastic scattering or neutrino interactions with Oxygen.} by the CCSN neutrino spectrum multiplied by the following kinematic factor: 
\begin{align}
w_{i} = \frac{\sigma(E_{p,i},\cos\theta_i)}{\int_{-1}^1\sigma(E_{p,i},\cos\theta')\;\mathrm{dcos}\theta'§}\frac{\mathrm{d}E_{p,i}}{\mathrm{d}E_{\nu,i}}.
\end{align}
Here, $E_{p,i}$ is the positron energy, $E_{\nu,i}$ the neutrino energy and $\sigma$ the cross section of the process considered. $\theta_i$ is the angle between the positron and neutrino directions. In this analysis, the neutrino direction is set to $\theta=\pi/3,$ and $\phi=4\pi/3$, corresponding to a CCSN located in the Galactic plane (most probable location for an observable supernova). The impact of the neutrino direction on the sensitivity of KM3NeT is however negligible. 

The CCSN simulation described above is used to assess the discriminating power of single-DOM observables, design a new single-DOM event selection strategy, and evaluate the sensitivity of KM3NeT to CCSNe. Simulated events are weighted according to the models used in the initial KM3NeT CCSN search~\cite{van2021km3net}.
\subsection{Radioactivity and atmospheric muon background simulations}
\label{sec:bkgmodels}
\subsubsection{Modeling radioactive decays}
\label{subsec:radioactivity}

Radioactive decays are simulated using the KM3NeT software package \texttt{OMGsim}~\cite{colomer2017detailed}. This  \texttt{GEANT4}-based package models the propagation of the decay products as well as the production and propagation of the Cherenkov photons. As for the CCSN signal, interaction vertices are simulated as being uniformly distributed within a sphere around a single DOM. A detailed technical description of this DOM is used to simulate photon propagation within its material and the final photoelectron production in the PMT photocathodes. The five dominant sources of radioactivity background in KM3NeT (using the respective decay rates in the deep seawater given in~\cite{albert2018long}) are the decays of $^{40}$K and $^{238}$U in water, and of $^{40}$K, $^{238}$U and $^{232}$Th in the DOM glass. The relative contributions of these decays to the total radioactivity background are given in Table~\ref{tab:decays}. For decays occurring in water, vertices are simulated within a radius of up to \SI{1}{\m} around the DOM. This radius has been shown to contain $99.5\%$ of the decays for events with multiplicity $4$ and above.

\begin{table}[]
    \centering
    \begin{tabular}{c|ccccc}
        \hline
         &  $^{40}$K water & $^{238}$U water  & $^{40}$K glass & $^{238}$U glass & $^{232}$Th glass\\
         \hline
         $M\geq 5$& $78\%$ & $6.1\%$&$3.5\%$ & $8.1\%$& $4.5\%$\\
         \hline
         $M\geq 7$& $62\%$ & $22$\%& $1.0\%$ & $6.5\%$& $9.1\%$\\
        \hline
    \end{tabular}
    \caption{Radioactivity background composition for multiplicities greater than or equal to $5$ (top row) and $7$ (bottom row). While $^{40}$K decays in water overwhelmingly dominate at low multiplicities, other decay processes can contribute by up to $\sim 10\%$ of the total background in the multiplicity range considered for CCSN studies.}
    \label{tab:decays}
\end{table}

\subsubsection{Modeling atmospheric muons}
\label{subsec:atmmu}


Muon energies, entry points, and momenta are simulated using \texttt{MUPAGE v1.5.1}~\cite{mupage1,mupage2}. The muon entry points are generated on the surface of a cylinder surrounding the detector. The radius and height of this cylinder are chosen such that muons passing up to \SI{140}{\m} (\SI{280}{\m}) away from ORCA (ARCA) are simulated, with contributions from other muons amounting to less than $0.5\%$ of the single-DOM event rate for multiplicities $6$ and above. The muon propagation and detection is simulated using the $\texttt{JSirene}$ software~\cite{jsirene1}, which models the Cherenkov light emission induced by muons using tabulated probability density functions instead of tracking each photon individually. The high computational efficiency of \texttt{JSirene} makes it possible to model muon backgrounds for multiple ARCA and ORCA geometries with minimal computational cost.
\printbibliography

@inproceedings{Chen:2016:XST:2939672.2939785,
 author = {Chen, Tianqi and Guestrin, Carlos},
 title = {{XGBoost}: A Scalable Tree Boosting System},
 booktitle = {Proceedings of the 22nd ACM SIGKDD International Conference on Knowledge Discovery and Data Mining},
 series = {KDD '16},
 year = {2016},
 isbn = {978-1-4503-4232-2},
 location = {San Francisco, California, USA},
 pages = {785--794},
 numpages = {10},
 url = {http://doi.acm.org/10.1145/2939672.2939785},
 doi = {10.1145/2939672.2939785},
 acmid = {2939785},
 publisher = {ACM},
 address = {New York, NY, USA},
}

@article{van2021km3net,
    author = "Aiello, S. and others",
    collaboration = "KM3NeT",
    title = "{The KM3NeT potential for the next core-collapse supernova observation with neutrinos}",
    % eprint = "2102.05977",
    % archivePrefix = "arXiv",
    % primaryClass = "astro-ph.HE",
    doi = "10.1140/epjc/s10052-021-09187-5",
    journal = "Eur. Phys. J. C",
    volume = "81",
    number = "5",
    pages = "445",
    year = "2021"
}

@article{aiello2022implementation,
  title={Implementation and first results of the KM3NeT real-time core-collapse supernova neutrino search},
  author={Aiello, Sebastiano and Albert, A and Alshamsi, M and Alves Garre, S and Aly, Z and Ambrosone, A and Ameli, F and Andre, M and Androulakis, G and Anghinolfi, M and others},
  journal={Eur. Phys. J. C},
  volume={82},
  number={4},
  pages={317},
  year={2022},
  publisher={Springer}
}

@article{priede2008potential,
  title={The potential influence of bioluminescence from marine animals on a deep-sea underwater neutrino telescope array in the Mediterranean Sea},
  author={Priede, Imants G and Jamieson, Alan and Heger, Amandine and Craig, Jessica and Zuur, Alain F},
  journal={Deep Sea Research Part I: Oceanographic Research Papers},
  volume={55},
  number={11},
  pages={1474--1483},
  year={2008},
  publisher={Elsevier}
}

@article{aiello2022km3net,
  title={The KM3NeT multi-PMT optical module},
  author={Aiello, Sebastiano and Albert, Arthur and Alshamsi, Mohammed and Garre, S Alves and Aly, Zineb and Ambrosone, Antonio and Ameli, Fabrizio and Andre, Michel and Androulakis, Giorgos and Anghinolfi, Marco and others},
  journal={Journal of Instrumentation},
  volume={17},
  number={07},
  pages={P07038},
  year={2022},
  publisher={IOP Publishing}
}

@article{walk2020neutrino,
  title={Neutrino emission characteristics of black hole formation in three-dimensional simulations of stellar collapse},
  author={Walk, Laurie and Tamborra, Irene and Janka, Hans-Thomas and Summa, Alexander and Kresse, Daniel},
  journal={Physical Review D},
  volume={101},
  number={12},
  pages={123013},
  year={2020},
  publisher={APS}
}

@article{colomer2017detailed,
  title={Detailed KM3NeT optical module simulation with Geant4 and supernova neutrino detection study},
  author={Colomer, Marta and Dornic, Damien and Kulikovskiy, Vladimir},
  journal={PoS (ICRC2017)},
  volume={983},
  year={2017}
}

@article{tsirigotis2011hou,
  title={HOU Reconstruction \& Simulation (HOURS): A complete simulation and reconstruction package for very large volume underwater neutrino telescopes},
  author={Tsirigotis, AG and Leisos, A and Tzamarias, SE},
  journal={Nucl. Instrum. Meth. A},
  volume={626},
  pages={S185--S187},
  year={2011},
  publisher={Elsevier}
}

@article{tamborra2014neutrino,
  title={Neutrino emission characteristics and detection opportunities based on three-dimensional supernova simulations},
  author={Tamborra, Irene and Raffelt, Georg and Hanke, Florian and Janka, Hans-Thomas and M{\"u}ller, Bernhard},
  journal={Physical Review D},
  volume={90},
  number={4},
  pages={045032},
  year={2014},
  publisher={APS}
}

@article{schramm1990new,
  title={New physics from supernova 1987A},
  author={Schramm, David N and Truran, James W},
  journal={Physics Reports},
  volume={189},
  number={2},
  pages={89--126},
  year={1990},
  publisher={Elsevier}
}

@article{goos2023searching,
%   title={Searching for Core-Collapse Supernova neutrinos at KM3NeT},
%   author={Goos, Isabel and El Hedri, Sonia and Donzaud, Corinne and Bendahman, Meriem and Collaboration, KM3NeT},
%   journal={Contributions of KM3NeT to ICRC2023},
%   pages={49},
%   year={2023}
% }

@article{suzuki2024neutrinos,
  title={Neutrinos from core-collapse supernova explosions},
  author={Suzuki, Hideyuki},
  journal={Progress of Theoretical and Experimental Physics},
  volume={2024},
  number={5},
  pages={05B101},
  year={2024},
  publisher={Oxford University Press}
}

@article{albert2018long,
  title={Long-term monitoring of the ANTARES optical module efficiencies using 40K decays in sea water},
  author={Albert, A and Andr{\'e}, Michel and Anghinolfi, Marco and Anton, G and Ardid, M and Aubert, J-J and Aublin, J and Avgitas, T and Baret, B and Barrios-Mart{\'\i}, J and others},
  journal={Eur. Phys. J. C},
  volume={78},
  number={8},
  pages={1--8},
  year={2018},
  publisher={Springer}
}

@article{rolke2005limits,
  title={Limits and confidence intervals in the presence of nuisance parameters},
  author={Rolke, Wolfgang A and L{\'o}pez, Angel M and Conrad, Jan},
  journal={Nucl. Instrum. Meth. A},
  volume={551},
  number={2-3},
  pages={493--503},
  year={2005},
  publisher={Elsevier}
}

@article{Adams:2013ana,
    author = "Adams, Scott M. and Kochanek, C. S. and Beacom, John F. and Vagins, Mark R. and Stanek, K. Z.",
    title = "{Observing the Next Galactic Supernova}",
    % eprint = "1306.0559",
    % archivePrefix = "arXiv",
    % primaryClass = "astro-ph.HE",
    doi = "10.1088/0004-637X/778/2/164",
    journal = "Astrophys. J.",
    volume = "778",
    pages = "164",
    year = "2013"
}

@article{mupage1,
    author = "Bazzotti, M. and Carminati, G. and Margiotta, A. and Spurio, M.",
    title = "{An update of the generator of atmospheric muons from parametric formulas (MUPAGE)}",
    doi = "10.1016/j.cpc.2009.12.017",
    journal = "Comput. Phys. Commun.",
    volume = "181",
    pages = "835--836",
    year = "2010"
}

@article{mupage2,
%     author = "Carminati, G. and Bazzotti, M. and Biagi, S. and Cecchini, S. and Chiarusi, T. and Margiotta, A. and Sioli, M. and Spurio, M.",
%     title = "{MUPAGE: a fast atmospheric MUon GEnerator for neutrino telescopes based on PArametric formulas}",
%     % eprint = "0907.5563",
%     % archivePrefix = "arXiv",
%     % primaryClass = "astro-ph.IM",
%     month = "7",
%     year = "2009"
% }

@article{dom_rates,
    author = "Adri\'an-Mart\'\i{}nez, S. and others",
    collaboration = "KM3NeT",
    title = "{Deep sea tests of a prototype of the KM3NeT digital optical module}",
    % eprint = "1405.0839",
    % archivePrefix = "arXiv",
    % primaryClass = "astro-ph.IM",
    doi = "10.1140/epjc/s10052-014-3056-3",
    journal = "Eur. Phys. J. C",
    volume = "74",
    number = "9",
    pages = "3056",
    year = "2014"
}

@article{radioactivity_simulation,
    author = "Herold, Bjorn",
    editor = "Anassontzis, E. G. and Rapidis, P. A. and Resvanis, L. K.",
    collaboration = "KM3NeT",
    title = "{Study of K-40-induced rates for a KM3NeT design option with multi-PMT optical modules}",
    doi = "10.1016/j.nima.2010.04.137",
    journal = "Nucl. Instrum. Meth. A",
    volume = "626-627",
    pages = "S234--S236",
    year = "2011"
}

@article{hirata1988observation,
  title={Observation in the Kamiokande-II detector of the neutrino burst from supernova SN1987A},
  author={Hirata, KS and Kajita, T and Koshiba, M and Nakahata, M and Oyama, Y and Sato, N and Suzuki, A and Takita, M and Totsuka, Y and Kifune, T and others},
  journal={Physical Review D},
  volume={38},
  number={2},
  pages={448},
  year={1988},
  publisher={APS}
}

@article{bionta1987observation,
  title={Observation of a neutrino burst in coincidence with supernova 1987A in the Large Magellanic Cloud},
  author={Bionta, RM and Blewitt, G and Bratton, CB and Casper, D and Ciocio, A and Claus, R and Cortez, B and Crouch, M and Dye, ST and Errede, S and others},
  journal={Physical Review Letters},
  volume={58},
  number={14},
  pages={1494},
  year={1987},
  publisher={APS}
}

@article{alekseev1987possible,
  title={Possible detection of a neutrino signal on 23 February 1987 at the Baksan underground scintillation telescope of the Institute of Nuclear Research},
  author={Alekseev, EN and Alekseeva, LN and Volchenko, VI and Krivosheina, IV},
  journal={JETP lett},
  volume={45},
  number={739},
  pages={1987},
  year={1987}
}

@article{walker1987making,
  title={Making the most of SN1987A},
  author={Walker, Terry P},
  journal={Nature},
  volume={330},
  number={6149},
  pages={609--610},
  year={1987},
  publisher={Nature Publishing Group UK London}
}

@article{horiuchi2018can,
  title={What can be learned from a future supernova neutrino detection?},
  author={Horiuchi, Shunsaku and Kneller, James P},
  journal={Journal of Physics G: Nuclear and Particle Physics},
  volume={45},
  number={4},
  pages={043002},
  year={2018},
  publisher={IOP Publishing}
}

@article{al2021snews,
  title={SNEWS 2.0: a next-generation supernova early warning system for multi-messenger astronomy},
  author={Al Kharusi, S and BenZvi, SY and Bobowski, JS and Bonivento, W and Brdar, V and Brunner, T and Caden, E and Clark, M and Coleiro, A and Colomer-Molla, M and others},
  journal={New Journal of Physics},
  volume={23},
  number={3},
  pages={031201},
  year={2021},
  publisher={IOP Publishing}
}

@article{coleiro2020combining,
  title={Combining neutrino experimental light-curves for pointing to the next galactic core-collapse supernova},
  author={Coleiro, Alexis and Molla, M Colomer and Dornic, Damien and Lincetto, Massimiliano and Kulikovskiy, Vladimir},
  journal={Eur. Phys. J. C},
  volume={80},
  number={9},
  pages={856},
  year={2020},
  publisher={Springer}
}

@article{ToT,
    author = "Bourlis, G. and Tsirigotis, A. G. and Tzamarias, S. E.",
    editor = "Hallewell, Gregory and Bradbury, Stella and Coyle, Paschal and Henry, Sylvain and Kappes, Alexander and Kooijman, P. and Piattelli, Paolo and Taiuti, Mauro",
    collaboration = "KM3NeT",
    title = "{Time over threshold electronics for an underwater neutrino telescope}",
    doi = "10.1016/j.nima.2008.12.081",
    journal = "Nucl. Instrum. Meth. A",
    volume = "602",
    pages = "129--132",
    year = "2009",
    note = "[Erratum: Nucl.Instrum.Meth.A 638, 206 (2011)]"
}

@article{jsirene1,
    author = "{d}e Jong, M. and van Campenhout, E.",
    title = "{The probability density function of the arrival time of {\v{C}}erenkov light}",
    eprint = "2305.19626",
    archivePrefix = "arXiv",
    primaryClass = "astro-ph.IM",
    month = "5",
    year = "2023"
}

@article{GEANT4,
    author = "Agostinelli, S. and others",
    collaboration = "GEANT4",
    title = "{GEANT4 - A Simulation Toolkit}",
    reportNumber = "SLAC-PUB-9350, FERMILAB-PUB-03-339, CERN-IT-2002-003",
    doi = "10.1016/S0168-9002(03)01368-8",
    journal = "Nucl. Instrum. Meth. A",
    volume = "506",
    pages = "250--303",
    year = "2003"
}

@article{KM3Net_LoI,
    author = "Adrian-Martinez, S. and others",
    collaboration = "KM3Net",
    title = "{Letter of intent for KM3NeT 2.0}",
    % eprint = "1601.07459",
    % archivePrefix = "arXiv",
    % primaryClass = "astro-ph.IM",
    doi = "10.1088/0954-3899/43/8/084001",
    journal = "J. Phys. G",
    volume = "43",
    number = "8",
    pages = "084001",
    year = "2016"
}

@article{KM3NeT:2024paj,
    author = "Aiello, S. and others",
    collaboration = "KM3NeT",
    title = "{Astronomy potential of KM3NeT/ARCA}",
    % eprint = "2402.08363",
    % archivePrefix = "arXiv",
    % primaryClass = "astro-ph.HE",
    doi = "10.1140/epjc/s10052-024-13137-2",
    journal = "Eur. Phys. J. C",
    volume = "84",
    number = "9",
    pages = "885",
    year = "2024"
}
\end{document}